\documentclass[aps,reprint,groupedaddress,showpacs,longbibliography,nofootinbib]{revtex4-2}
\usepackage{amssymb}
\usepackage{amsmath}
\usepackage[breaklinks,colorlinks,citecolor=green,urlcolor=blue,linkcolor=red]{hyperref}
\usepackage{graphicx}
\usepackage{epstopdf}
\usepackage{diagbox}
\usepackage{float}
\usepackage{subfigure}
\usepackage{multirow}
\usepackage{CJK}
\usepackage{tikz}
\usepackage{enumerate}
\usepackage{makecell}
\usepackage{enumitem}

\usepackage[markup=default]{changes}
\newcommand{\stkout}[1]{\ifmmode\text{\sout{\ensuremath{#1}}}\else\sout{#1}\fi}
\setdeletedmarkup{\stkout{#1}}

\begin{document}

\newcommand{\RN}[1]{\uppercase\expandafter{\romannumeral#1}}
\newcommand{\CPV}{$C\!P$V}
\newcommand{\CPA}{$C\!P$A}


\title{\boldmath Full analysis of CP violation induced by the decay angular correlations in  four-body cascade decays of heavy hadrons}


\author{Zhen-Hua Zhang}
\email{zhangzh@usc.edu.cn}
\author{Jian-Yu Yang}
\email{yangjy@stu.usc.edu.cn}
\affiliation{School of Nuclear Science and Technology, University of South China, Hengyang, 421001, Hunan, China}
\author{Xin-Heng Guo}
\email{xhguo@bnu.edu.cn}
\affiliation{School of Physical Science and Technology, Kunming University, Kunming 650214, China}


\date{\today}
\begin{abstract}
The violation of the charge-parity (CP) transformation symmetry, which although has been observed in plenty of pure meson decay processes, was only confirmed just very recently by the LHCb collaboration in the four-body decay of the heavy baryon $\Lambda_b^0$, $\Lambda_b^0\to p K^- \pi^+ \pi ^-$, through a comparison of the decay branching ratio with that of the CP-conjugate process.
However, the detailed dynamics behind this CP asymmetry is obviously far from clear.
In this paper, we propose a formalism for the full analysis of the decay angular correlations in four-body cascade decays of heavy hadrons which can provide more information about the CP violation in these decays.
To illustrate this, we apply the decay angular correlation analysis of CP violation to another four-body decay channel that involve baryons, $B^0\to p\bar{p}K^+\pi^-$, which has also been investigated by the LHCb collaboration with no evidence of CP violation being found.
Surprisingly, based on a simple assumption on the statistical errors, and with the event yield extracted inversely from the published data of LHCb, we obtain non-zero CP asymmetries of about $10\%$ corresponding to the decay angular correlations, which are considerably larger than the CPA asymmetries observed in the $\Lambda_b^0\to p K^- \pi^+ \pi ^-$ channel. 
We suggest our experimental colleagues to perform full decay angular correlation analyses of CP violation in four-body decays of heavy hadrons, including the above two decay channels.
\end{abstract}

\maketitle

{\it Introduction---}
CP violation (CPV) is a very important concept both in particle physics and cosmology.
Within the Standard Model (SM) of particle physics, CPV is described by the weak phase in the Cabibbo-Kobayashi-Maskawa (CKM) matrix \cite{Kobayashi:1973fv,Cabibbo:1963yz}.
CPV is also an essential ingredient for the generation of the Baryon Asymmetry of the observable Universe \cite{Sakharov:1967dj},
though the CKM mechanism in the SM is far from enough to account for the observed net baryon density of the Universe \cite{Trodden:1998ym,Riotto:1999yt,Morrissey:2012db,Planck:2018vyg}. 

CPV has already been observed in the pure mesonic decays of $K$, $B$, and $D$ mesons \cite{Christenson:1964fg,LHCb:2019hro,BaBar:2001ags,Belle:2001zzw,LHCb:2013syl,Workman:2022ynf}.
However, the searching of CPV in processes with baryons involved was long missing. 
There are two typical kinds of decay processes involving baryons, which include baryon-to-baryon and meson-to-baryon-anti-baryon-pair transitions, respectively.
It is only until very recently that the LHCb collaboration made a breakthrough in the CPV searching in decay processes involving baryons. 
They observed a non-zero CP asymmetry (CPA) in $\Lambda_b^0\to p K^- \pi^+ \pi ^-$ \cite{LHCb:2025ray}, confirming CPV in this baryon-to-baryon four-body decay channel.
However, CPV in the other type of decays which involve the meson-to-baryon-anti-baryon-pair transition has not been observed despite some existing evidence \cite{LHCb:2014nix}.

The threshold effect makes the branching ratios of the meson-to-baryon-anti-baryon-pair transitions of the type $M\to \mathbf{B}\overline{\mathbf{B}'} X$ be enhanced comparing with the type $M\to \mathbf{B}\overline{\mathbf{B}'}$, where $M$ is the mother meson, $\mathbf{B}\overline{\mathbf{B}'}$ is the baryon-anti-baryon pair, $X$ represents other particles, typically, one or more mesons \cite{Hou:2000bz}, or even more exotic scenarios \cite{LHCb:2025uxu}.
Therefore, the decays of the type $M\to \mathbf{B}\overline{\mathbf{B}'} X$ have advantage over $M\to \mathbf{B}\overline{\mathbf{B}'}$ due to higher statistics of the data, which is very important for CPV study.
Moreover, the presence of particles, $X$, leads to more complex phase space with more freedom to construct CPV observables corresponding to the angular correlations of the particles involved in the decays.

CPV corresponding to the decay angular correlations has been long studied both theoretically and experimentally.
Perhaps the most famous one is the one induced by the Triple-Product Asymmetry (TPA) in four-body decays of heavy hadrons, {\it i.e.}, the distribution asymmetry corresponding to the signature of the triple-products of the momenta and/or spins of the particle involved \cite{Donoghue:1987wu,Valencia:1988it,Dunietz:1990cj,Golowich:1988ig,Kayser:1989vw,Bensalem:2000hq,Bensalem:2002pz,Bensalem:2002ys,Durieux:2015zwa,Gronau:2015gha,Durieux:2016nqr,Shi:2019vus,Wang:2022fih} .
One typical example lies in the decay of a heavy meson into two vector mesons, $M_Q\to V_a V_b$, with the two vector mesons $V_a$ and $V_b$ respectively decaying into two pseudo-scalar ones.
The TPA induced CPAs (TP-CPAs) have not been observed yet, even though experimental investigations have been performed in various decay channels. 
Although a lot attentions have been paid to CPV corresponding to the decay angular correlations both theoretically and experimentally, such as the aforementioned TPA, a general and complete analysis of the CPV corresponding to the decay angular correlations is hardly seen in literature \cite{Dunietz:1990cj}.
One analysis of such kind was presented in Ref. \cite{Durieux:2015zwa} for four-body decays of a heavy meson with all the four particles in the final state being spinless pseudo-scalar mesons. 

In this paper, we propose to search CPV through a full angular correlation analysis in four-body decays of heavy hadrons.
When applying our formalism to the four-body decay process $B^0\to p\bar{p}K^+\pi^-$, we find surprisingly that clear evidences of CPV emerge in some of the angular correlations. 

{\it Decay angular correlations and the corresponding CPAs in four-body cascade decays---}
We consider the four-body decays of a heavy hadron, $H_Q\to 1234$, where the invariant masses of the 12 and 34 systems respectively lie closely around resonances $a_k$ and $b_m$, so that the decay is dominated by the cascade one $H_Q\to a_k(\to12) b_m(\to34)$, and the off-shell effects are small and the $q^2$ dependence of the form factors, with $q$ being a proper energy-momentum transfer, is mild.
The subscripts $k$ and $m$ ($k,m=1,2,\cdots$) indicate that there may be more than one resonances with similar masses in each decay branch.
Besides resonances, there could also be some smooth contributions with definite spin-parity quantum numbers, which are also included if not explicit mentioned.
In general, there will be interferences between decay amplitudes mediated by different $a_k$'s and/or $b_m$'s \footnote{In order for the interference between two resonances to happen, the difference between the masses of $a_k$ ($b_m$) for different $k$ ($m$) should be small enough so that it is comparable with their decay widths.}.
The inclusion of these interferences is essential, which will also turn out to be of great importance, to a full analysis of the decay angular correlations.  
In what follows, we will collectively denote the possible multiple resonances as $a$ and $b$ respectively.

\begin{figure}[b]\centering
\begin{tikzpicture}[xshift=-0.5,yshift=0.5,xscale=.5,yscale=.5]
\tikzstyle{every node}=[scale=0.75]
\begin{scope}[rotate=30]

\draw[stealth-,thick](5.4,1.2)--(6,0);   \node[rotate=-30]at (5.4,1.4){2};    
\draw[- stealth,thick](3,0)--(3.6,1.2);  \node[rotate=-30]at (3.5,1.4){2};    
\draw[- stealth,thick](-3,0)--(-3.7,1.4);   \node[rotate=-30]at (-3.9,1.6){3};
\draw[stealth -,thick](-5.3,1.4)--(-6,0);   \node[rotate=-30]at (-5.2,1.6){3};
\draw[->] (-6.4,0) arc (180:64.5:0.4); \node[rotate=15]at (-6.5,0.6){$\theta_b$}; 

\draw [rotate=-15,xslant =0.268, draw=none,fill = blue!20,blue!20,fill opacity=0.7] (-8,-2.5) rectangle (-5,0);  
\draw [rotate=-15,xslant =0.268, draw=none,fill = blue!20,blue!20,fill opacity=0.7] (-3.5,-2.5) rectangle (3.5,0); 
\draw [rotate=-15,xslant =0.268, draw=none,fill = blue!20,blue!20,fill opacity=0.7] (5,-2.5) rectangle (8,0); 

\draw[rotate=-15,xslant =0.268,blue!20] (-8,-2.5)--(8,-2.5); 
\draw[stealth -,thick](-3,0)--(0,0); 

\draw [rotate=-30,xslant =1, draw=none, fill = red!20,red!20,fill opacity=0.7] (-2,-1) rectangle (2,1);  

\draw [rotate=-15,xslant =0.268, draw=none,fill = blue!20,blue!20,fill opacity=0.7] (-8,0) rectangle (-5,2.5); 
\draw [rotate=-15,xslant =0.268, draw=none,fill = blue!20,blue!20,fill opacity=0.7] (-3.5,0) rectangle (3.5,2.5); 
\draw [rotate=-15,xslant =0.268, draw=none,fill = blue!20,blue!20,fill opacity=0.7] (5,0) rectangle (8,2.5); 

\draw[rotate=-15,xslant =0.268,blue!20] (-8,2.5)--(8,2.5); 

\draw[latex -,dotted, thick](-8,0)--(-3,0); 
\draw[- latex,dotted, thick](3,0)--(8,0); 

\draw[- latex,dashed,rotate=-30](0,0)--(0,1.15470054); 
\draw[- latex,dashed,rotate=105](0,0)--(0,1.5); 
\draw[- latex,dashed,rotate=-120](0,0)--(0,2); 

\draw[->] (0.4,0) arc (0:80:0.3); 

\node[rotate=15]at (0.1,-0.5){$H_Q$}; 
\node[rotate=15] at (2.7,-0.4) {$a$};
\node[rotate=15] at (-2.7,-0.4) {$b$};

\filldraw[black] (6,0) circle (1pt); 
\draw[- stealth,thick](6,0)--(6.6,-1.2); 

\draw[- stealth,thick](3,0)--(3.9,-1.2); 

\filldraw[black] (-6,0) circle (1pt); 
\draw[- stealth,thick](-6,0)--(-6.7,-1.4); 

\draw[- stealth,thick](-3,0)--(-4.2,-1.4); 

\draw[->] (7.2,0) arc (180:-150:0.15 and 0.35); 
\draw[->] (-7.3,0) arc (-180:150:0.15 and 0.35); 
\draw[->] (6.4,0) arc (0:-62.5:0.4); 

\filldraw[black] (0,0) circle (2pt); 

\draw[- stealth,thick](0,0)--(3,0); 

\node[rotate=15]at (0.6,1.4){$\hat{P}_{H_Q}$};
\node[rotate=15]at (1.1,0.5){$\theta_{{\scriptscriptstyle H_Q}}$}; 
\node[rotate=15]at (6.7,-0.4){$\theta_a$}; 
\node[rotate=15]at (7.7,-0.6){$\phi_a$};
\node[rotate=15]at (-7.5,0.65){$\phi_b$};
\node[rotate=15]at (6.6,-1.4){1}; 
\node[rotate=15]at (4,-1.4){1}; 
\node[rotate=15]at (-6.9,-1.6){4};
\node[rotate=15]at (-4.3,-1.6){4};
 
\node[rotate=15]at (8.3,0){$z_a$};  \node[rotate=15]at (-8.3,0){$z_b$};
 
\draw[rotate=-15,xslant =0.268,black,dotted] (-8,0)--(8,0); 

\node[gray,rotate=15] at (-1,-2) {$H_Q$ c.m. frame};
\node[gray,rotate=15] at (5.8,-2.5) {$a$ helicity frame};
\node[gray,rotate=15] at (-5.8,2.5) {$b$ helicity frame};

\end{scope}
\end{tikzpicture}
\caption{Illustration of the five angles defined in the main text for the decay $H_Q\to a(\to 12) b(\to 34)$. 
}\label{fig:Kinematics}
\end{figure}
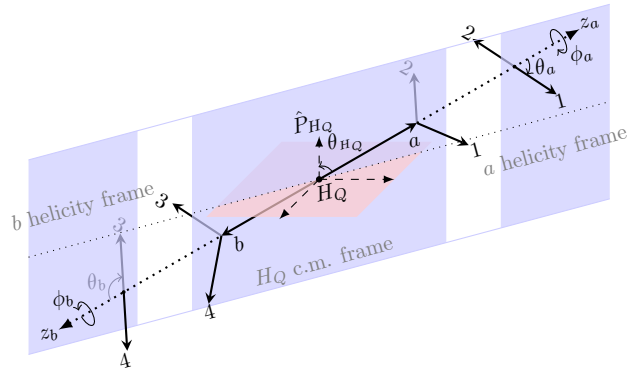

We adopt the generalized Cabibbo-Maksymowicz parametrization of the phase space to describe the decays \cite{Cabibbo:1965zzb}. 
For polarized hadrons, we need seven kinematical variables, which are two invariant masses and five angles. 
The invariant masses of 12 and 34 will be denoted as $m_{12}$ and $m_{34}$, respectively.
The five angles are: the helicity angle of $a$ in the c.m. frame of $H_Q$, which will be denoted as $\theta_{H_Q}$, the helicity angle of 1 (3) in the c.m. frame of ${a(b)}$, which will be denoted as $\theta_{a(b)}$, 
and the azimuthal angle between the $a\to 12$ ($b\to 34$) decay plane and the spin-quantization direction of $H_Q$, $\vec{P}_{H_q}$, which will be denoted as $\phi_{a(b)}$.
These five angles are illustrated in Fig. \ref{fig:Kinematics}.
We further introduce $\phi\equiv\phi_a-\phi_b$ and $\varphi\equiv\phi_a+\phi_b$ to replace $\phi_a$ and $\phi_b$.
One reason for this replacement is that when $H_Q$ is unpolarized, $\phi$ will be simply unphysical.
The other reason is that the angle $\varphi$, which is in fact the angle between the decay plane of $a$ and that of $b$, is independent of the polarization of $H_Q$.

In this paper, we will focus mainly on CP violation for unpolarized $H_Q$ 
\footnote{The discussion in the main text also applies to the cases when $H_Q$ is a polarized hadron but with the polarization information being integrated out.}.
Some discussions on polarized $H_Q$ are presented in Appendix.
For unpolarized $H_Q$, the decay amplitude squared (DAS) for $H_Q \to a(\to 12) b(\to 34)$ can be expressed as 
\begin{equation}\label{eq:AngDisCom}
	{\left| \mathcal{A}\right|^2} \propto \sum_{j_a,j_b,\sigma}\left[ \Re(\gamma^{j_aj_b}_{\sigma})\Psi^{j_aj_b}_{\sigma} -\Im(\gamma^{j_aj_b}_{\sigma})\Phi^{j_aj_b}_{\sigma} \right],
\end{equation}
where $\gamma^{j_aj_b}_{\sigma}$ is the dynamical factor which form is presented in Appendix, $\Re$ and $\Im$ mean taking the real and imaginary parts, $\Psi^{j_aj_b}_{\sigma}$ and $\Phi^{j_aj_b}_{\sigma}$ are the kinematical factors, which respectively take the froms
\begin{equation}\label{Eq:Phi}
	\Phi_{\sigma}^{j_aj_b}=d^{j_a}_{\sigma,0}(\theta_a)d^{j_b}_{\sigma,0}(\theta_b)\sin\sigma\varphi,
\end{equation}
and 
\begin{equation}\label{Eq:Psi}
	\Psi_{\sigma}^{j_aj_b}=d^{j_a}_{\sigma,0}(\theta_a)d^{j_b}_{\sigma,0}(\theta_b)\cos\sigma\varphi,
\end{equation}
where $d$'s are the Wigner-d-matrices, $j_x$ ($x=a,b$) is the angular quantum number of the sum $\mathbf{S}_x+\mathbf{S}_{x'}$, with $\mathbf{S}_{x^{(\prime)}}$ the spin operator of $x^{(\prime)}$ ($x$ and $x'$ comes respectively from the decay amplitude $\mathcal{A}$ and its complex conjugate in Eq. (\ref{eq:AngDisCom})), $\sigma$ is defined in Appendix. 
The explicit forms of the first few kinematical factors are presented in Appendix.


Once $a_k$ and $b_m$ are selected, all the possible decay angular correlations are settled. 
There are respectively two constraints for the integers
 $j_a$ and $j_b$.
The first one is the triangular inequality
\begin{equation}\label{eq:so3con}
|s_{a_k}-s_{a_{k'}}| \leqslant j_a \leqslant s_{a_k}+s_{a_{k'}},
\end{equation}
which must satisfied by $j_a$
for a given pair of $a_k$ and $a_{k'}$.
The second one comes from the parity symmetry in the strong decay $a\to 12$, which provides the constraints for $j_a$ through 
\begin{equation}\label{eq:paritycon}
(-)^{j_a}=\Pi_{a_k}\Pi_{a_{k'}},
\end{equation}
where $\Pi$'s represent the parities of the corresponding particles.
Letting $a_k$ and $a_{k'}$ run over all the allowed possibilities, we obtain all the allowed values of $j_a$.
There are also two similar constraints for $j_b$ which we will not show explicitly here.
The index $\sigma$ 
is constrained by $-j_a\leqslant\sigma \leqslant j_a$ and $-j_b\leqslant\sigma \leqslant j_b$, 
which implies that the number of the independent decay angular correlations is $C^{{j_a}{j_b}}=2\min({j_a},{j_b})+1$ for a given pair of $j_a$ and $j_b$, where we have introduced a matrix $C$ for convenience. 
Once the allowed values for $j_a$ and $j_b$ are determined, the number of the independent angular correlations for a given pair of sets $\{s_{a_m}\}$ and $\{s_{b_k}\}$ can be obtained via
\begin{equation}\label{Eq:NumAngCo}
	\mathcal{N}_{\{s_{a_m}\}\{s_{b_k}\}}=\sum_{ {j_a},{j_b}} C^{{j_a}{j_b}}.
\end{equation}

Once all the allowed decay angular correlations are written down, we can construct the corresponding  CPV observables as
$A_{CP}^{\Re({\gamma}^{j_aj_b}_{\sigma})}= \rho_{\Re({\gamma}^{j_aj_b}_{\sigma})}\frac{\Re({\gamma}^{j_aj_b}_{\sigma})-\Re(\overline{\gamma}^{j_aj_b}_{\sigma})}{{\gamma}^{00}_{0}+\overline{\gamma}^{00}_{0}}$, and 
$A_{CP}^{\Im({\gamma}^{j_aj_b}_{\sigma})}=\rho_{\Im({\gamma}^{j_aj_b}_{\sigma})} \frac{\Im({\gamma}^{j_aj_b}_{\sigma})+\Im(\overline{\gamma}^{j_aj_b}_{\sigma})}{{\gamma}^{00}_{0}+\overline{\gamma}^{00}_{0}}$, where $\overline{\gamma}^{j_aj_b}_{\sigma}$ is the CP-conjugate of ${\gamma}^{j_aj_b}_{\sigma}$, $\rho_{\Re({\gamma}^{j_aj_b}_{\sigma})}$ and $\rho_{\Im({\gamma}^{j_aj_b}_{\sigma})}$ are constants whose explicit values are left open here \footnote{We thank Prof. Haiyong Wang (School of Mathematics and Statistics, Huazhong University of Science and Technology) on the discussion on this point.}.
In the Standard Model, the CP violation, which manifests as differences between these decay-angular distributions in $B^0$ and $\overline{B^0}$ decays, are induced by the weak CKM phase and strong rescattering phases of the helicity amplitudes in the interference terms.
One advantage of the set of CPA observables $\{A_{CP}^{\Re({\gamma}^{j_aj_b}_{\sigma})},A_{CP}^{\Im({\gamma}^{j_aj_b}_{\sigma})}\}$ is that the contributions are theoretically clean because  (1) different kinematic factors are orthogonal, and (2) the interference between a pair of amplitudes with a certain spin-parity and helicity quantum numbers can only contribute to one of the above angular-correlated CPA observables \footnote{There may be contamination from other four-body decay topologies, such as $H_Q\to c 1$, with $c\to d 2$, and $d\to 34 $. However, it can be understood that this decay topology only affects a small fraction of bins. Consequently, its contribution to the angular-correlated CPAs considered in this Letter will be small.}.
Alternatively, one can also define the CPV observables as
\begin{equation}\label{Eq:YACP}
	A_{CP}^{\mathcal{Y}^{j_aj_b}_\sigma}
	\!\!\!\equiv\!\! \frac{\left(N_{\mathcal{Y}^{j_aj_b}_\sigma>0}\!-\!N_{\mathcal{Y}^{j_aj_b}_\sigma<0}\right)
		\!-\!
		\left(N_{\overline{\mathcal{Y}}^{j_aj_b}_\sigma>0}\!-\!N_{\overline{\mathcal{Y}}^{j_aj_b}_\sigma<0}\right)}{N+\overline{N}},
\end{equation}
where $\mathcal{Y}$ can be either $\Phi$ or $\Psi$, $\overline{\mathcal{Y}}^{j_aj_b}_\sigma$ is the CP-conjugate of ${\mathcal{Y}}^{j_aj_b}_\sigma$, $\overline{\mathcal{Y}}^{j_aj_b}_\sigma(\bar{\theta}_a,\bar{\theta}_b,\bar{\varphi})\equiv {\delta_\mathcal{Y}} \mathcal{Y}(\bar{\theta}_a,\bar{\theta}_b,\bar{\varphi})$, with $\delta_{\Psi}=+1$, and $\delta_{\Phi}=-1$.
In principle, the two sets of CPA observables, $\{A_{CP}^{\Re({\gamma}^{j_aj_b}_{\sigma})},A_{CP}^{\Im({\gamma}^{j_aj_b}_{\sigma})}\}$ and $\{A^{\mathcal{Y}^{j_aj_b}_\sigma}\}$, are equivalent to each other and can be expressed by each other. 
For example, $A_{CP}^{\mathcal{Y}^{j_aj_b}_\sigma}=\sum_{j_a'j_b'\sigma'} \lambda^{j_aj_bj_a'j_b'}_{\sigma\sigma'}
  A_{CP}^{\mathcal{C}^{j_a'j_b'}_{\sigma'}}$,
  where $\mathcal{C}^{j_a'j_b'}_{\sigma'}$ is a collective abbreviation of $\Re({\gamma}^{j_aj_b}_{\sigma})$ and $\Im({\gamma}^{j_aj_b}_{\sigma})$, $\lambda^{j_aj_bj_a'j_b'}_{\sigma\sigma'}\propto\int \text{sgn}(\mathcal{Y}^{j_aj_b}_\sigma)\mathcal{Y}^{j_a'j_b'}_{\sigma'} d\Omega$, 
  with $d\Omega$ being the phase space integral element over $\theta_a$, $\theta_b$, and $\varphi$. 
  From the expression of $\lambda^{j_aj_bj_a'j_b'}_{\sigma\sigma'}$, it is easily seen that it is only when $j_a'=j_a$, $j_b'=j_b$, and $\sigma'=\sigma$ that there is no cancellation between different parts of the integral, 
  indicating that $A^{\Psi^{j_aj_b}_\sigma}_{CP}$ ($A^{\Phi^{j_aj_b}_\sigma}_{CP}$) gets its main contribution from $A_{CP}^{\Re({\gamma}^{j_aj_b}_{\sigma})}$ ($A_{CP}^{\Im({\gamma}^{j_aj_b}_{\sigma})}$)
\footnote{The normalization constants $\rho_{\Re({\gamma}^{j_aj_b}_{\sigma})}$ and $\rho_{\Im({\gamma}^{j_aj_b}_{\sigma})}$ may be chosen such that $\rho_{\mathcal{C}^{j_aj_bj_aj_b}_{\sigma\sigma}}=1$. It can be seen that $\big|\rho_{\mathcal{C}^{j_aj_bj_a'j_b'}_{\sigma\sigma'}}\big|$ will be considerably smaller that 1 if either $j_a\neq j_a'$, $j_b\neq j_b'$, or $\sigma\neq \sigma'$ is fulfilled. See Ref. \cite{Qi:2025zna} for more information about the normalization problem.}.

We have presented CPAs corresponding to all the angular correlations among the three kinematical angles $\theta_a$, $\theta_b$, and $\varphi$ for the four-body decays $H_Q\to a(\to12)b(\to34)$, which serve as the foundation for the CP-violation analysis discussed below.
However, to complete the discussion, there is still one important issue missing.
As was discussed, we also take into account the interferences between different intermediate resonances.
But when considering the corresponding CP violation, only one part of the interference term, either the real or the imaginary part, contributes to the CPV observable. 
It is possible sometime to retrieve the other part of the interference term. 
We will come to this point in the application part.

{\it Application to $B^0\to p\bar{p}K^+\pi^-$---}
A very interesting example to illustrate our formalism is $B^0\to p\bar{p}K^+\pi^-$, which is a type of meson-to-baryon-anti-baryon-pair transition with no evidence of CPV was found before \cite{LHCb:2022orj}. 
It has a large contribution from the $K^\ast(892)$ resonance in the low $K^+\pi^-$ invariant mass region, and experiences a threshold enhancement in the low ${p\bar{p}}$ invariant mass region, as can be seen clearly from the first two rows of the subfigures in Fig. 5 in Ref. \cite{LHCb:2022orj}. 
The two subfigures in the third row of Fig. 5 show a clear forward-backward asymmetry (FBA) in $\cos\theta_{p\bar{p}}$, which is a common characteristic of the baryon-anti-baryon production in heavy meson decay processes \cite{Belle:2003pwf,Suzuki:2006nn,Geng:2006wz}.
There are also FBAs in $\cos\theta_{K\pi}$ of about 10\% and 12\% for the $B^0$ and $\overline{B^0}$ case, respectively, which can be read off from the two subfigures of the fourth row, though it is hard to be recognized by eye. 
Although there is no clear hint of TPA corresponding to $\sin\varphi$ in the last row of Fig. 5, it does show interestingly a very clear left-right asymmetry (LRA) corresponding to $\cos\varphi$. 

\begin{table*}
	\centering
	\begin{tabular}{c||c|c|c|c|c|c|c|c|c|c|c|c|c||c|c|c|c|c|c}
		\hline\hline
	\diagbox[height=25pt,innerwidth=34pt]{CPAs}{${\mathcal{Y}}^{j_aj_b}_{\sigma}$} & $\Psi^{00}_0$ & $\Psi^{01}_0$ & $\Psi^{10}_0$ & $\Psi^{11}_0$ & $\Psi^{11}_1$ & $\Phi^{11}_1$ & $\Psi^{12}_1$ & $\Phi^{12}_1$ & $\Psi^{21}_1$ & $\Phi^{21}_1$ & $\Psi^{22}_1$ & $\Phi^{22}_1$ & $\Phi^{22}_2$ & $\Psi^{02}_0$ & $\Psi^{20}_0$ & $\Psi^{12}_0$ & $\Psi^{21}_0$ & $\Psi^{22}_0$ & $\Psi^{22}_2$ \\
		\hline
		${A}_{CP}^{\mathcal{Y}^{j_aj_b}_{\sigma}}$ & 5.8 & 8.5 & -5.6 & -2.8 & 2.5 & -4.0 & {\bf 9.2} & -0.8 & -1.7 & -5.6 & 0.4 & -2.0 & -3.4 &&&&&&\\
		\hline
		$\tilde{A}_{CP}^{\mathcal{Y}^{j_aj_b}_{\sigma}}$ &  \diagbox[dir=SW]{ }{ } & {\bf 10.7} &  \diagbox[dir=SW]{ }{ } & {\bf  10.0} & -0.0 & -0.5 &  \diagbox[dir=SW]{ }{ } &  \diagbox[dir=SW]{ }{ } & -4.3 & -2.4 & \diagbox[dir=SW]{ }{ } & \diagbox[dir=SW]{ }{ } & \diagbox[dir=SW]{ }{} & \diagbox[dir=SW]{ }{ } & \diagbox[dir=SW]{ }{ } & \diagbox[dir=SW]{ }{ } &&\diagbox[dir=SW]{ }{ }&\diagbox[dir=SW]{ }{ }\\
		\hline\hline
	\end{tabular}
	\caption{Decay angular correlation CPAs in unit of $\%$ calculated with the event yields extracted from the data in Ref. \cite{LHCb:2022orj}. The cells left blank are observalbes which are not available with the currently extracted data, hence are expected to be measured by LHCb in the future. The cells with slash symbols do not need to be measured, as are explained in the main text. Statistical and systematical errors, which are not presented in this table, are discussed in the main texts and Appendix.}\label{Tab:ACPX}
\end{table*}	

When combining the above observations, we conclude:
\begin{itemize}[leftmargin=0pt, itemindent=1em,]
  \item The FBAs in the $p\bar{p}$ and $K^+\pi^-$ systems are caused mainly by the angular correlations $\Psi^{10}_0$ and $\Psi^{01}_0$, respectively.
One should note that the angular correlations like $\Psi^{11}_0$, $\Psi^{12}_0$, and $\Psi^{21}_0$ can not contribute to the aforementioned FBAs since the integration of these angular correlations with respect to $\cos\theta_{a}$ and $\cos\theta_{b}$ vanishes.
  \item The LRA corresponding to $\cos\varphi$ is induced mainly by $\Phi^{11}_1$.
Again, the contributions of other angular correlations which can produce LRAs such as $\Phi^{21}_1$, $\Phi^{12}_1$, and $\Phi^{22}_1$, vanish once $\cos\theta_a$ and $\cos\theta_b$ are integrated out.
\end{itemize}
The presence of the angular correlations $\Psi^{10}_0$, $\Psi^{01}_0$, and $\Phi^{11}_1$ shows that $j_a$ and $j_b$ can take values $j_a,j_b=0, 1$.
The presence of the odd number values of $j_a$ and $j_b$ indicates that there are interferences between contributions with opposite parities, as can be seen from Eq. (\ref{eq:paritycon}).
This indicates strongly that we have (at least for the simplest scenario) contributions (resonances and/or smooth non-resonances) with spin-parities ${s_{a_1}}^{\!\!\Pi_{a_1}}={0^\pm}$, ${s_{a_2}}^{\!\!\Pi_{a_2}}={1^\mp}$, and ${s_{b_1}}^{\!\!\Pi_{b_1}}=1^-$, ${s_{b_2}}^{\!\!\Pi_{b_2}}=0^+$, where $b_1$ is recognized as $K^\ast(892)$ whose signal is clearly shown \footnote{It is also possible that the $p\bar{p}$ system gets contributions from $1^-$ and $1^+$ channels.}.
The presence of the spin-1 contributions in both of the $p\bar{p}$ and $K^+\pi^-$ systems indicates that the maximum values for $j_a$ and $j_b$ can be 2, as can be deduced from Eq. (\ref{eq:so3con}).
According to the above analysis, there are 19 angular correlations, as can be seen obviously from Eq. (\ref{Eq:NumAngCo}). 

Moreover, as was mentioned above, after integrating $m_{34}^2$ (which will be denoted as $m_{K\pi}^2$ for the current situation) around the mass-squared of $K^\ast(892)$, only one part of the interference term between $K^\ast(892)$ and the scalar ($s_{a_2}=0$) part survives, either the real or the imaginary part, leaving the other part integrated out from the CPV observables constructed above. 
To be more explicit, the interference term is proportional to 
$\Re(\frac{\lambda}{\mathcal{I}_{K^\ast}})\propto \frac{\mathcal{I}_{K^\ast}^\ast\lambda}{|\mathcal{I}_{K^\ast}|^2}\propto (m_{K\pi}^2-m^2_{K^\ast})\Re(\lambda)+(m_{K^\ast}\Gamma_{K^\ast})\Im(\lambda)$, if the $s_{a_2}=0$ scalar part is smooth or a broad resonance, where $\lambda$ is the interference term.
One can see clearly that if $m_{K\pi}^2$ is integrated out around $m_{K^\ast}^2$, the term $\Re(\lambda)$ will simply have little contribution.
But CPV could also lies in $\Re(\lambda)$.
The integrated-out part $\Re(\lambda)$ can be retrieved by introducing an extra sign factor $\text{sgn}_{34}\equiv\text{sgn}(m_{34}^2-m_{K^\ast}^2)$ inside the integral \cite{Durieux:2015zwa,Qi:2024zau}.
This will introduce extra CPV observables, which can be defined as
		$\tilde{A}_{CP}^{\mathcal{Y}^{j_aj_b}_\sigma}
	\!\!\equiv\!\!  \big[(N_{\text{sgn}_{34}\mathcal{Y}^{j_aj_b}_\sigma>0} - N_{\text{sgn}_{34}\mathcal{Y}^{j_aj_b}_\sigma<0})
		-
		(N_{\text{sgn}_{34}\overline{\mathcal{Y}}^{j_aj_b}_\sigma>0} - N_{\text{sgn}_{34}\overline{\mathcal{Y}}^{j_aj_b}_\sigma<0})\big]/(N+\overline{N})$.
Note that we only need to introduce this kind of CPV observables for decay angular correlations where the interference between the resonance $K^\ast(892)$ and the $0^+$ scalar part appears, which corresponds to $j_b=1$.
This corresponds to 7 more angular-correlated CPA observables, as can be seen from TABLE \ref{Tab:AnCo} in Appendix.
It seems that we do not need to introduce such kind of CPV observables in the $p\bar{p}$ system, because both $0^\pm$ and $1^\mp$ contributions seem to be from the threshold enhancement effects rather than resonances.

With the TPAs presented in Ref. \cite{LHCb:2022orj}, we can extract the event yields. Thanks to the binning approach adopted by LHCb in Ref. \cite{LHCb:2022orj}, i.e., binning according to the sign of $\cos\theta_a$, $\cos\theta_b$, $\cos\varphi$, and $\sin\varphi$, plus two alternative binning schemes---which are referred as Schemes A and B in Ref. \cite{LHCb:2022orj}---according to whether or not the signs of  $m_{K\pi}^2-m_{K^\ast}^2$ are taken into account, we are able to calculate 19 out of the 26 angular-correlated CPAs, which are listed in TABLE \ref{Tab:ACPX}.
Surprisingly, we obtain quite large CPAs corresponding to some of these angular correlations in the aforementioned decay channel, {\it i.e.}, $\tilde{A}_{CP}^{\Psi^{01}_{0}}=(10.7\pm2.05\pm0.21)\%$, $\tilde{A}_{CP}^{\Psi^{11}_{0}}=(10.0\pm2.05\pm0.21)\%$, and $A_{CP}^{\Psi^{12}_{1}}=(9.2\pm2.07\pm0.21)\%$, respectively, where the first (second) error is the statistical (systematic) one, which will be explained in the next paragraph \footnote{It should be pointed out that the CPAs of ten percent level are quite large CPAs, comparing with, for example, the recently observed CP asymmetries in $\Lambda_b^0\to p K^- \pi^+ \pi^-$, although such large CPAs are understandable within the SM, given proper strong phase difference between the interference terms.}.
The first one corresponds to the contribution from the interference between $K^\ast(892)$ and the ${s_{b_2}}^{\!\!\Pi_{b_2}}=0^+$. 
The second one, $\tilde{A}_{CP}^{\Psi^{11}_{0}}$, which was referred as the two-fold FBA induced CPA in Ref. \cite{Zhang:2022iye}, corresponds to the interference terms $\mathcal{A}_{a_1b_1}\mathcal{A}_{a_2b_2}^\ast$ and $\mathcal{A}_{a_1b_2}\mathcal{A}_{a_2b_1}^\ast$, with $\mathcal{A}_{a_kb_m}$ being the amplitude mediated by $a_k$ and $b_m$ (we have omitted all other indices in the decay amplitudes).
The fact of $\tilde{A}_{CP}^{\Psi^{11}_{0}}$ being non-zero indicates that the amplitudes corresponding to the helicity $\pm1$ of $K^\ast(892)$ are non-zero, since this angular correlation corresponds to the dynamical term which contains $\mathcal{F}^{(0^\pm, 1^-)}_{0,0}\mathcal{F}^{(1^\mp, 1^-)\ast}_{+1,+1}$ and $\mathcal{F}^{(1^\mp, 1^-)}_{-1,-1}\mathcal{F}^{(0^\pm, 1^-)\ast}_{0,0}$, where the spin-parities of $a$ and $b$ are emphasized in the superscripts of the helicity amplitudes, hence a further investigation of the polarization of $K^\ast(892)$ in this channel is urgent \cite{Belle:2008zkc}.
The last one corresponds to the interferences between ${s_{a_1}}^{\!\!\Pi_{a_1}}={0^\pm}$ and ${s_{a_1}}^{\!\!\Pi_{a_1}}={1^\mp}$ in the $p\bar{p}$ system. 
The observation of non-zero CPA corresponding to $A_{CP}^{\Psi^{12}_{1}}$ will indicate both the $0^\pm$ and $1^\mp$ contributions are important to the threshold enhancement.
Interestingly, all the three angular correlations obtain contribution from interfering terms which contain $\mathcal{F}^{(0^\pm, 1^-)}_{0,0}$.
None of the above three CPAs correspond to TPAs.

As can be deduced from TABLEs \ref{Tab:B0Data} and \ref{Tab:B0barData} in Appendix, the extracted total event yields for $B^0$ and $\overline{B^0}$ are $N_{B^0}^{\text{Scheme A}}=2972$, $N_{\overline{B^0}}^{\text{Scheme A}}=2646$, and $N_{B^0}^{\text{Scheme B}}=2985$, $N_{\overline{B^0}}^{\text{Scheme B}}=2719$, for Schemes A and B, respectively. 
Consequently, the statistical errors of these CPA observables in the first and the second rows of TABLE \ref{Tab:ACPX}, which correspond to the total event yields of Schemes A and B respectively, can be calculated to be 2.07 and 2.05, respectively, according to $\sigma_{\text{stat.}}=k/\sqrt{N_{B^0}+N_{\overline{B^0}}}$, where $k$ is the scale factor introduced in Appendix. As mentioned in the analysis of LHCb in Ref. \cite{LHCb:2022orj}, the systematic errors, which come from the detector resolution, fitting procedure, alternative fit, and mass resolution, are estimated to be about 9\% of the statistical ones.

The large significance of non-zeroness of the angular correlated CPAs benefit from the substantially small statistical errors, as can been clearly seen from TABLE \ref{Tab:ACPX}.
This is a very important feature of the angular correlation analysis, comparing with the regional CPA analysis, which is widely adopted in experimental studies in multi-body decays of heavy hadrons. 

Usually, one needs first to find out the possible resonances, in order to list all the possible decay angular correlations.
By exhaustive measurements of the CPAs corresponding to all the possible decay angular correlations, one can find where CPV is located and further find out the dynamics behind it according to Eq. (\ref{eq:AngDisCom}).  
It should be pointed out that our method also works if one does not know the possible resonances.
One just needs to measure the CPAs corresponding to the decay angular correlations  from $\mathcal{Y}^{00}_0$, $\mathcal{Y}^{10}_0$, $\mathcal{Y}^{01}_0$, $\mathcal{Y}^{11}_0$, $\mathcal{Y}^{11}_1$, $\cdots$, step by step. 

The results in TABLE \ref{Tab:ACPX} depend on the assumption we have made when extracting the signal yields, especially on the universality of the scaling factor $k$ which is introduced in Appendix.
Hence, rather than as a strong evidence of CPV, we would conservatively regard the large significance in some of the CPA observables in TABLE \ref{Tab:ACPX} as a persuasive suggestion to our experimental colleagues to reexamine this decay channel via a full angular correlation analysis of CPAs, in which the aforementioned assumption-dependence will be eliminated.
Moreover, although CPV was observed in the decay $\Lambda_b^0\to p K^- \pi^+ \pi^-$, the concrete reasons (such as the interference between what kinds of amplitudes) for this CPV are still not clear.
Hence a full analysis of CPV corresponding to decay angular correlations is necessary.

{\it Summary and outlook---}
In summary, we propose a formalism for investigating CPV via the full decay angular correlation analysis in four-body cascade decays of heavy hadrons.  
The interferences between decay paths mediated by different resonances are taken into account in the formalism, which turns out to be very important for CPV studies in four-body decays.

The analysis of the decay  $B^0\to p\bar{p}K^+\pi^-$ based on the data of LHCb collaboration shows that there are several interesting advantages of studying CPV via a full angular-correlation analysis in four-body decays of heavy hadrons.
First of all, it is model-independent.
Secondly, it appears to be efficient, comparing with other methods such as the amplitude analysis and the energy test methods \cite{Williams:2011cd,LHCb:2016qbq,Parkes:2016yie}.
Thirdly, this method connect the CPV observables and the dynamics inducing the CPV in a more transparent way.
Last but not least, it make use of all the data in the region considered, hence has a much smaller statistical uncertainty comparing with the CPV observables defined in part of the region which has been widely studied experimentally in multi-body decays of heavy hadrons.
In fact, the proposed method is the only one which possesses all the four features, comparing with other widely used methods, including the aforementioned amplitude analysis method, energy test method, and regional CPA analysis in different parts of the phase space.
A comparison of the proposed method with other widely adopted methods is presented in TABEL. \ref{Tab:Comp}.
We suggest our experimental colleagues to apply this method in the CPV studies of four-body decays of heavy hadrons.

\begin{table}
	\centering
	\caption{ Comparison of the features of different methods of CPV analysis} \label{Tab:Comp}
	\begin{tabular}{|c|c|c|c|c|}
		\hline
		\diagbox[height=35pt,innerwidth=35pt]{methods}{features} & \makecell[l]{statistical\\ significance} & \makecell[l]{inferring \\dynamics} & \makecell[l]{model\\-independent} & efficiency \\
		\hline
		regional CPA &$\times$&$\times$&$\checkmark$&$\checkmark$\\ \hline
		\makecell[l]{amplitude\\ analysis}&$\checkmark$&$\checkmark$&$\times$&$\times$\\ \hline
		energy test&$\checkmark$& $\times$ &$\checkmark$&$\times$\\ \hline
		this work & $\checkmark$ &$\checkmark$&$\checkmark$&$\checkmark$\\
		\hline
	\end{tabular}
\end{table}

{\it Acknowledgments---}
We thank Dr. Longke Li, Dr. Yanxi Zhang, Prof. Yu-Kuo Hsiao and Prof. Kuang-Ta Chao for valuable discussions. 
This work was supported by National Natural Science Foundation of China under Grants Nos. 12475096, 12275024, and 12192261, Scientific Research Fund of Hunan Provincial Education Department under Grants No. 22A0319.

\bibliography{zzhbib}

\begin{thebibliography}{42}%
\makeatletter
\providecommand \@ifxundefined [1]{%
 \@ifx{#1\undefined}
}%
\providecommand \@ifnum [1]{%
 \ifnum #1\expandafter \@firstoftwo
 \else \expandafter \@secondoftwo
 \fi
}%
\providecommand \@ifx [1]{%
 \ifx #1\expandafter \@firstoftwo
 \else \expandafter \@secondoftwo
 \fi
}%
\providecommand \natexlab [1]{#1}%
\providecommand \enquote  [1]{``#1''}%
\providecommand \bibnamefont  [1]{#1}%
\providecommand \bibfnamefont [1]{#1}%
\providecommand \citenamefont [1]{#1}%
\providecommand \href@noop [0]{\@secondoftwo}%
\providecommand \href [0]{\begingroup \@sanitize@url \@href}%
\providecommand \@href[1]{\@@startlink{#1}\@@href}%
\providecommand \@@href[1]{\endgroup#1\@@endlink}%
\providecommand \@sanitize@url [0]{\catcode `\\12\catcode `\$12\catcode
  `\&12\catcode `\#12\catcode `\^12\catcode `\_12\catcode `\%12\relax}%
\providecommand \@@startlink[1]{}%
\providecommand \@@endlink[0]{}%
\providecommand \url  [0]{\begingroup\@sanitize@url \@url }%
\providecommand \@url [1]{\endgroup\@href {#1}{\urlprefix }}%
\providecommand \urlprefix  [0]{URL }%
\providecommand \Eprint [0]{\href }%
\providecommand \doibase [0]{https://doi.org/}%
\providecommand \selectlanguage [0]{\@gobble}%
\providecommand \bibinfo  [0]{\@secondoftwo}%
\providecommand \bibfield  [0]{\@secondoftwo}%
\providecommand \translation [1]{[#1]}%
\providecommand \BibitemOpen [0]{}%
\providecommand \bibitemStop [0]{}%
\providecommand \bibitemNoStop [0]{.\EOS\space}%
\providecommand \EOS [0]{\spacefactor3000\relax}%
\providecommand \BibitemShut  [1]{\csname bibitem#1\endcsname}%
\let\auto@bib@innerbib\@empty
\bibitem [{\citenamefont {Kobayashi}\ and\ \citenamefont
  {Maskawa}(1973)}]{Kobayashi:1973fv}%
  \BibitemOpen
  \bibfield  {author} {\bibinfo {author} {\bibfnamefont {M.}~\bibnamefont
  {Kobayashi}}\ and\ \bibinfo {author} {\bibfnamefont {T.}~\bibnamefont
  {Maskawa}},\ }\bibfield  {title} {\bibinfo {title} {{CP Violation in the
  Renormalizable Theory of Weak Interaction}},\ }\href
  {https://doi.org/10.1143/PTP.49.652} {\bibfield  {journal} {\bibinfo
  {journal} {Prog. Theor. Phys.}\ }\textbf {\bibinfo {volume} {49}},\ \bibinfo
  {pages} {652} (\bibinfo {year} {1973})}\BibitemShut {NoStop}%
\bibitem [{\citenamefont {Cabibbo}(1963)}]{Cabibbo:1963yz}%
  \BibitemOpen
  \bibfield  {author} {\bibinfo {author} {\bibfnamefont {N.}~\bibnamefont
  {Cabibbo}},\ }\bibfield  {title} {\bibinfo {title} {{Unitary Symmetry and
  Leptonic Decays}},\ }\href {https://doi.org/10.1103/PhysRevLett.10.531}
  {\bibfield  {journal} {\bibinfo  {journal} {Phys. Rev. Lett.}\ }\textbf
  {\bibinfo {volume} {10}},\ \bibinfo {pages} {531} (\bibinfo {year}
  {1963})}\BibitemShut {NoStop}%
\bibitem [{\citenamefont {Sakharov}(1967)}]{Sakharov:1967dj}%
  \BibitemOpen
  \bibfield  {author} {\bibinfo {author} {\bibfnamefont {A.~D.}\ \bibnamefont
  {Sakharov}},\ }\bibfield  {title} {\bibinfo {title} {{Violation of CP
  Invariance, C asymmetry, and baryon asymmetry of the universe}},\ }\href
  {https://doi.org/10.1070/PU1991v034n05ABEH002497} {\bibfield  {journal}
  {\bibinfo  {journal} {Pisma Zh. Eksp. Teor. Fiz.}\ }\textbf {\bibinfo
  {volume} {5}},\ \bibinfo {pages} {32} (\bibinfo {year} {1967})}\BibitemShut
  {NoStop}%
\bibitem [{\citenamefont {Trodden}(1999)}]{Trodden:1998ym}%
  \BibitemOpen
  \bibfield  {author} {\bibinfo {author} {\bibfnamefont {M.}~\bibnamefont
  {Trodden}},\ }\bibfield  {title} {\bibinfo {title} {{Electroweak
  baryogenesis}},\ }\href {https://doi.org/10.1103/RevModPhys.71.1463}
  {\bibfield  {journal} {\bibinfo  {journal} {Rev. Mod. Phys.}\ }\textbf
  {\bibinfo {volume} {71}},\ \bibinfo {pages} {1463} (\bibinfo {year}
  {1999})},\ \Eprint {https://arxiv.org/abs/hep-ph/9803479}
  {arXiv:hep-ph/9803479} \BibitemShut {NoStop}%
\bibitem [{\citenamefont {Riotto}\ and\ \citenamefont
  {Trodden}(1999)}]{Riotto:1999yt}%
  \BibitemOpen
  \bibfield  {author} {\bibinfo {author} {\bibfnamefont {A.}~\bibnamefont
  {Riotto}}\ and\ \bibinfo {author} {\bibfnamefont {M.}~\bibnamefont
  {Trodden}},\ }\bibfield  {title} {\bibinfo {title} {{Recent progress in
  baryogenesis}},\ }\href {https://doi.org/10.1146/annurev.nucl.49.1.35}
  {\bibfield  {journal} {\bibinfo  {journal} {Ann. Rev. Nucl. Part. Sci.}\
  }\textbf {\bibinfo {volume} {49}},\ \bibinfo {pages} {35} (\bibinfo {year}
  {1999})},\ \Eprint {https://arxiv.org/abs/hep-ph/9901362}
  {arXiv:hep-ph/9901362} \BibitemShut {NoStop}%
\bibitem [{\citenamefont {Morrissey}\ and\ \citenamefont
  {Ramsey-Musolf}(2012)}]{Morrissey:2012db}%
  \BibitemOpen
  \bibfield  {author} {\bibinfo {author} {\bibfnamefont {D.~E.}\ \bibnamefont
  {Morrissey}}\ and\ \bibinfo {author} {\bibfnamefont {M.~J.}\ \bibnamefont
  {Ramsey-Musolf}},\ }\bibfield  {title} {\bibinfo {title} {{Electroweak
  baryogenesis}},\ }\href {https://doi.org/10.1088/1367-2630/14/12/125003}
  {\bibfield  {journal} {\bibinfo  {journal} {New J. Phys.}\ }\textbf {\bibinfo
  {volume} {14}},\ \bibinfo {pages} {125003} (\bibinfo {year} {2012})},\
  \Eprint {https://arxiv.org/abs/1206.2942} {arXiv:1206.2942 [hep-ph]}
  \BibitemShut {NoStop}%
\bibitem [{\citenamefont {Aghanim}\ \emph {et~al.}(2020)\citenamefont {Aghanim}
  \emph {et~al.}}]{Planck:2018vyg}%
  \BibitemOpen
  \bibfield  {author} {\bibinfo {author} {\bibfnamefont {N.}~\bibnamefont
  {Aghanim}} \emph {et~al.} (\bibinfo {collaboration} {Planck}),\ }\bibfield
  {title} {\bibinfo {title} {{Planck 2018 results. VI. Cosmological
  parameters}},\ }\href {https://doi.org/10.1051/0004-6361/201833910}
  {\bibfield  {journal} {\bibinfo  {journal} {Astron. Astrophys.}\ }\textbf
  {\bibinfo {volume} {641}},\ \bibinfo {pages} {A6} (\bibinfo {year} {2020})},\
  \bibinfo {note} {[Erratum: Astron.Astrophys. 652, C4 (2021)]},\ \Eprint
  {https://arxiv.org/abs/1807.06209} {arXiv:1807.06209 [astro-ph.CO]}
  \BibitemShut {NoStop}%
\bibitem [{\citenamefont {Christenson}\ \emph {et~al.}(1964)\citenamefont
  {Christenson}, \citenamefont {Cronin}, \citenamefont {Fitch},\ and\
  \citenamefont {Turlay}}]{Christenson:1964fg}%
  \BibitemOpen
  \bibfield  {author} {\bibinfo {author} {\bibfnamefont {J.~H.}\ \bibnamefont
  {Christenson}}, \bibinfo {author} {\bibfnamefont {J.~W.}\ \bibnamefont
  {Cronin}}, \bibinfo {author} {\bibfnamefont {V.~L.}\ \bibnamefont {Fitch}},\
  and\ \bibinfo {author} {\bibfnamefont {R.}~\bibnamefont {Turlay}},\
  }\bibfield  {title} {\bibinfo {title} {{Evidence for the $2\pi$ Decay of the
  $K_2^0$ Meson}},\ }\href {https://doi.org/10.1103/PhysRevLett.13.138}
  {\bibfield  {journal} {\bibinfo  {journal} {Phys. Rev. Lett.}\ }\textbf
  {\bibinfo {volume} {13}},\ \bibinfo {pages} {138} (\bibinfo {year}
  {1964})}\BibitemShut {NoStop}%
\bibitem [{\citenamefont {Aaij}\ \emph {et~al.}(2019)\citenamefont {Aaij} \emph
  {et~al.}}]{LHCb:2019hro}%
  \BibitemOpen
  \bibfield  {author} {\bibinfo {author} {\bibfnamefont {R.}~\bibnamefont
  {Aaij}} \emph {et~al.} (\bibinfo {collaboration} {LHCb}),\ }\bibfield
  {title} {\bibinfo {title} {{Observation of CP Violation in Charm Decays}},\
  }\href {https://doi.org/10.1103/PhysRevLett.122.211803} {\bibfield  {journal}
  {\bibinfo  {journal} {Phys. Rev. Lett.}\ }\textbf {\bibinfo {volume} {122}},\
  \bibinfo {pages} {211803} (\bibinfo {year} {2019})},\ \Eprint
  {https://arxiv.org/abs/1903.08726} {arXiv:1903.08726 [hep-ex]} \BibitemShut
  {NoStop}%
\bibitem [{\citenamefont {Aubert}\ \emph {et~al.}(2001)\citenamefont {Aubert}
  \emph {et~al.}}]{BaBar:2001ags}%
  \BibitemOpen
  \bibfield  {author} {\bibinfo {author} {\bibfnamefont {B.}~\bibnamefont
  {Aubert}} \emph {et~al.} (\bibinfo {collaboration} {BaBar}),\ }\bibfield
  {title} {\bibinfo {title} {{Measurement of CP violating asymmetries in $B^0$
  decays to CP eigenstates}},\ }\href
  {https://doi.org/10.1103/PhysRevLett.86.2515} {\bibfield  {journal} {\bibinfo
   {journal} {Phys. Rev. Lett.}\ }\textbf {\bibinfo {volume} {86}},\ \bibinfo
  {pages} {2515} (\bibinfo {year} {2001})},\ \Eprint
  {https://arxiv.org/abs/hep-ex/0102030} {arXiv:hep-ex/0102030} \BibitemShut
  {NoStop}%
\bibitem [{\citenamefont {Abe}\ \emph {et~al.}(2001)\citenamefont {Abe} \emph
  {et~al.}}]{Belle:2001zzw}%
  \BibitemOpen
  \bibfield  {author} {\bibinfo {author} {\bibfnamefont {K.}~\bibnamefont
  {Abe}} \emph {et~al.} (\bibinfo {collaboration} {Belle}),\ }\bibfield
  {title} {\bibinfo {title} {{Observation of large CP violation in the neutral
  $B$ meson system}},\ }\href {https://doi.org/10.1103/PhysRevLett.87.091802}
  {\bibfield  {journal} {\bibinfo  {journal} {Phys. Rev. Lett.}\ }\textbf
  {\bibinfo {volume} {87}},\ \bibinfo {pages} {091802} (\bibinfo {year}
  {2001})},\ \Eprint {https://arxiv.org/abs/hep-ex/0107061}
  {arXiv:hep-ex/0107061} \BibitemShut {NoStop}%
\bibitem [{\citenamefont {Aaij}\ \emph {et~al.}(2013)\citenamefont {Aaij} \emph
  {et~al.}}]{LHCb:2013syl}%
  \BibitemOpen
  \bibfield  {author} {\bibinfo {author} {\bibfnamefont {R.}~\bibnamefont
  {Aaij}} \emph {et~al.} (\bibinfo {collaboration} {LHCb}),\ }\bibfield
  {title} {\bibinfo {title} {{First observation of $CP$ violation in the decays
  of $B^0_s$ mesons}},\ }\href {https://doi.org/10.1103/PhysRevLett.110.221601}
  {\bibfield  {journal} {\bibinfo  {journal} {Phys. Rev. Lett.}\ }\textbf
  {\bibinfo {volume} {110}},\ \bibinfo {pages} {221601} (\bibinfo {year}
  {2013})},\ \Eprint {https://arxiv.org/abs/1304.6173} {arXiv:1304.6173
  [hep-ex]} \BibitemShut {NoStop}%
\bibitem [{\citenamefont {Workman}(2022)}]{Workman:2022ynf}%
  \BibitemOpen
  \bibfield  {author} {\bibinfo {author} {\bibfnamefont {R.~L.}\ \bibnamefont
  {Workman}} (\bibinfo {collaboration} {Particle Data Group}),\ }\bibfield
  {title} {\bibinfo {title} {{Review of Particle Physics}},\ }\href
  {https://doi.org/10.1093/ptep/ptac097} {\bibfield  {journal} {\bibinfo
  {journal} {PTEP}\ }\textbf {\bibinfo {volume} {2022}},\ \bibinfo {pages}
  {083C01} (\bibinfo {year} {2022})}\BibitemShut {NoStop}%
\bibitem [{\citenamefont {Aaij}\ \emph
  {et~al.}(2025{\natexlab{a}})\citenamefont {Aaij} \emph
  {et~al.}}]{LHCb:2025ray}%
  \BibitemOpen
  \bibfield  {author} {\bibinfo {author} {\bibfnamefont {R.}~\bibnamefont
  {Aaij}} \emph {et~al.} (\bibinfo {collaboration} {LHCb}),\ }\bibfield
  {title} {\bibinfo {title} {{Observation of charge{\textendash}parity symmetry
  breaking in baryon decays}},\ }\href
  {https://doi.org/10.1038/s41586-025-09119-3} {\bibfield  {journal} {\bibinfo
  {journal} {Nature}\ }\textbf {\bibinfo {volume} {643}},\ \bibinfo {pages}
  {1223} (\bibinfo {year} {2025}{\natexlab{a}})},\ \Eprint
  {https://arxiv.org/abs/2503.16954} {arXiv:2503.16954 [hep-ex]} \BibitemShut
  {NoStop}%
\bibitem [{\citenamefont {Aaij}\ \emph {et~al.}(2014)\citenamefont {Aaij} \emph
  {et~al.}}]{LHCb:2014nix}%
  \BibitemOpen
  \bibfield  {author} {\bibinfo {author} {\bibfnamefont {R.}~\bibnamefont
  {Aaij}} \emph {et~al.} (\bibinfo {collaboration} {LHCb}),\ }\bibfield
  {title} {\bibinfo {title} {{Evidence for CP Violation in $B^+\to p \overline
  p K^+$ Decays}},\ }\href {https://doi.org/10.1103/PhysRevLett.113.141801}
  {\bibfield  {journal} {\bibinfo  {journal} {Phys. Rev. Lett.}\ }\textbf
  {\bibinfo {volume} {113}},\ \bibinfo {pages} {141801} (\bibinfo {year}
  {2014})},\ \Eprint {https://arxiv.org/abs/1407.5907} {arXiv:1407.5907
  [hep-ex]} \BibitemShut {NoStop}%
\bibitem [{\citenamefont {Hou}\ and\ \citenamefont {Soni}(2001)}]{Hou:2000bz}%
  \BibitemOpen
  \bibfield  {author} {\bibinfo {author} {\bibfnamefont {W.-S.}\ \bibnamefont
  {Hou}}\ and\ \bibinfo {author} {\bibfnamefont {A.}~\bibnamefont {Soni}},\
  }\bibfield  {title} {\bibinfo {title} {{Pathways to rare baryonic B
  decays}},\ }\href {https://doi.org/10.1103/PhysRevLett.86.4247} {\bibfield
  {journal} {\bibinfo  {journal} {Phys. Rev. Lett.}\ }\textbf {\bibinfo
  {volume} {86}},\ \bibinfo {pages} {4247} (\bibinfo {year} {2001})},\ \Eprint
  {https://arxiv.org/abs/hep-ph/0008079} {arXiv:hep-ph/0008079} \BibitemShut
  {NoStop}%
\bibitem [{\citenamefont {Aaij}\ \emph
  {et~al.}(2025{\natexlab{b}})\citenamefont {Aaij} \emph
  {et~al.}}]{LHCb:2025uxu}%
  \BibitemOpen
  \bibfield  {author} {\bibinfo {author} {\bibfnamefont {R.}~\bibnamefont
  {Aaij}} \emph {et~al.} (\bibinfo {collaboration} {LHCb}),\ }\bibfield
  {title} {\bibinfo {title} {{First observation of the charmless baryonic decay
  $B^+\to\bar{\Lambda} p \bar{p} p$}},\ }\href@noop {} {\  (\bibinfo {year}
  {2025}{\natexlab{b}})},\ \Eprint {https://arxiv.org/abs/2508.16259}
  {arXiv:2508.16259 [hep-ex]} \BibitemShut {NoStop}%
\bibitem [{\citenamefont {Donoghue}\ \emph {et~al.}(1987)\citenamefont
  {Donoghue}, \citenamefont {Holstein},\ and\ \citenamefont
  {Valencia}}]{Donoghue:1987wu}%
  \BibitemOpen
  \bibfield  {author} {\bibinfo {author} {\bibfnamefont {J.~F.}\ \bibnamefont
  {Donoghue}}, \bibinfo {author} {\bibfnamefont {B.~R.}\ \bibnamefont
  {Holstein}},\ and\ \bibinfo {author} {\bibfnamefont {G.}~\bibnamefont
  {Valencia}},\ }\bibfield  {title} {\bibinfo {title} {{Survey of Present and
  Future Tests of {CP} Violation}},\ }\href
  {https://doi.org/10.1142/S0217751X87000144} {\bibfield  {journal} {\bibinfo
  {journal} {Int. J. Mod. Phys. A}\ }\textbf {\bibinfo {volume} {2}},\ \bibinfo
  {pages} {319} (\bibinfo {year} {1987})}\BibitemShut {NoStop}%
\bibitem [{\citenamefont {Valencia}(1989)}]{Valencia:1988it}%
  \BibitemOpen
  \bibfield  {author} {\bibinfo {author} {\bibfnamefont {G.}~\bibnamefont
  {Valencia}},\ }\bibfield  {title} {\bibinfo {title} {{Angular Correlations in
  the Decay $B \to V V$ and {CP} Violation}},\ }\href
  {https://doi.org/10.1103/PhysRevD.39.3339} {\bibfield  {journal} {\bibinfo
  {journal} {Phys. Rev. D}\ }\textbf {\bibinfo {volume} {39}},\ \bibinfo
  {pages} {3339} (\bibinfo {year} {1989})}\BibitemShut {NoStop}%
\bibitem [{\citenamefont {Dunietz}\ \emph {et~al.}(1991)\citenamefont
  {Dunietz}, \citenamefont {Quinn}, \citenamefont {Snyder}, \citenamefont
  {Toki},\ and\ \citenamefont {Lipkin}}]{Dunietz:1990cj}%
  \BibitemOpen
  \bibfield  {author} {\bibinfo {author} {\bibfnamefont {I.}~\bibnamefont
  {Dunietz}}, \bibinfo {author} {\bibfnamefont {H.~R.}\ \bibnamefont {Quinn}},
  \bibinfo {author} {\bibfnamefont {A.}~\bibnamefont {Snyder}}, \bibinfo
  {author} {\bibfnamefont {W.}~\bibnamefont {Toki}},\ and\ \bibinfo {author}
  {\bibfnamefont {H.~J.}\ \bibnamefont {Lipkin}},\ }\bibfield  {title}
  {\bibinfo {title} {{How to extract CP violating asymmetries from angular
  correlations}},\ }\href {https://doi.org/10.1103/PhysRevD.43.2193} {\bibfield
   {journal} {\bibinfo  {journal} {Phys. Rev. D}\ }\textbf {\bibinfo {volume}
  {43}},\ \bibinfo {pages} {2193} (\bibinfo {year} {1991})}\BibitemShut
  {NoStop}%
\bibitem [{\citenamefont {Golowich}\ and\ \citenamefont
  {Valencia}(1989)}]{Golowich:1988ig}%
  \BibitemOpen
  \bibfield  {author} {\bibinfo {author} {\bibfnamefont {E.}~\bibnamefont
  {Golowich}}\ and\ \bibinfo {author} {\bibfnamefont {G.}~\bibnamefont
  {Valencia}},\ }\bibfield  {title} {\bibinfo {title} {{Triple Product
  Correlations in Semileptonic $B^\pm$ Decays}},\ }\href
  {https://doi.org/10.1103/PhysRevD.40.112} {\bibfield  {journal} {\bibinfo
  {journal} {Phys. Rev. D}\ }\textbf {\bibinfo {volume} {40}},\ \bibinfo
  {pages} {112} (\bibinfo {year} {1989})}\BibitemShut {NoStop}%
\bibitem [{\citenamefont {Kayser}(1990)}]{Kayser:1989vw}%
  \BibitemOpen
  \bibfield  {author} {\bibinfo {author} {\bibfnamefont {B.}~\bibnamefont
  {Kayser}},\ }\bibfield  {title} {\bibinfo {title} {{Kinematically Nontrivial
  {CP} Violation in Beauty Decay}},\ }\href
  {https://doi.org/10.1016/0920-5632(90)90113-9} {\bibfield  {journal}
  {\bibinfo  {journal} {Nucl. Phys. B Proc. Suppl.}\ }\textbf {\bibinfo
  {volume} {13}},\ \bibinfo {pages} {487} (\bibinfo {year} {1990})}\BibitemShut
  {NoStop}%
\bibitem [{\citenamefont {Bensalem}\ and\ \citenamefont
  {London}(2001)}]{Bensalem:2000hq}%
  \BibitemOpen
  \bibfield  {author} {\bibinfo {author} {\bibfnamefont {W.}~\bibnamefont
  {Bensalem}}\ and\ \bibinfo {author} {\bibfnamefont {D.}~\bibnamefont
  {London}},\ }\bibfield  {title} {\bibinfo {title} {{$T$ odd triple product
  correlations in hadronic $b$ decays}},\ }\href
  {https://doi.org/10.1103/PhysRevD.64.116003} {\bibfield  {journal} {\bibinfo
  {journal} {Phys. Rev. D}\ }\textbf {\bibinfo {volume} {64}},\ \bibinfo
  {pages} {116003} (\bibinfo {year} {2001})},\ \Eprint
  {https://arxiv.org/abs/hep-ph/0005018} {arXiv:hep-ph/0005018} \BibitemShut
  {NoStop}%
\bibitem [{\citenamefont {Bensalem}\ \emph
  {et~al.}(2002{\natexlab{a}})\citenamefont {Bensalem}, \citenamefont {Datta},\
  and\ \citenamefont {London}}]{Bensalem:2002pz}%
  \BibitemOpen
  \bibfield  {author} {\bibinfo {author} {\bibfnamefont {W.}~\bibnamefont
  {Bensalem}}, \bibinfo {author} {\bibfnamefont {A.}~\bibnamefont {Datta}},\
  and\ \bibinfo {author} {\bibfnamefont {D.}~\bibnamefont {London}},\
  }\bibfield  {title} {\bibinfo {title} {{T-violating triple-product
  correlations in charmless $\Lambda_b$ decays}},\ }\href
  {https://doi.org/10.1016/S0370-2693(02)02028-2} {\bibfield  {journal}
  {\bibinfo  {journal} {Phys. Lett. B}\ }\textbf {\bibinfo {volume} {538}},\
  \bibinfo {pages} {309} (\bibinfo {year} {2002}{\natexlab{a}})},\ \Eprint
  {https://arxiv.org/abs/hep-ph/0205009} {arXiv:hep-ph/0205009} \BibitemShut
  {NoStop}%
\bibitem [{\citenamefont {Bensalem}\ \emph
  {et~al.}(2002{\natexlab{b}})\citenamefont {Bensalem}, \citenamefont {Datta},\
  and\ \citenamefont {London}}]{Bensalem:2002ys}%
  \BibitemOpen
  \bibfield  {author} {\bibinfo {author} {\bibfnamefont {W.}~\bibnamefont
  {Bensalem}}, \bibinfo {author} {\bibfnamefont {A.}~\bibnamefont {Datta}},\
  and\ \bibinfo {author} {\bibfnamefont {D.}~\bibnamefont {London}},\
  }\bibfield  {title} {\bibinfo {title} {{New-physics effects on triple-product
  correlations in $\Lambda_b$ decays}},\ }\href
  {https://doi.org/10.1103/PhysRevD.66.094004} {\bibfield  {journal} {\bibinfo
  {journal} {Phys. Rev. D}\ }\textbf {\bibinfo {volume} {66}},\ \bibinfo
  {pages} {094004} (\bibinfo {year} {2002}{\natexlab{b}})},\ \Eprint
  {https://arxiv.org/abs/hep-ph/0208054} {arXiv:hep-ph/0208054} \BibitemShut
  {NoStop}%
\bibitem [{\citenamefont {Durieux}\ and\ \citenamefont
  {Grossman}(2015)}]{Durieux:2015zwa}%
  \BibitemOpen
  \bibfield  {author} {\bibinfo {author} {\bibfnamefont {G.}~\bibnamefont
  {Durieux}}\ and\ \bibinfo {author} {\bibfnamefont {Y.}~\bibnamefont
  {Grossman}},\ }\bibfield  {title} {\bibinfo {title} {{Probing CP violation
  systematically in differential distributions}},\ }\href
  {https://doi.org/10.1103/PhysRevD.92.076013} {\bibfield  {journal} {\bibinfo
  {journal} {Phys. Rev. D}\ }\textbf {\bibinfo {volume} {92}},\ \bibinfo
  {pages} {076013} (\bibinfo {year} {2015})},\ \Eprint
  {https://arxiv.org/abs/1508.03054} {arXiv:1508.03054 [hep-ph]} \BibitemShut
  {NoStop}%
\bibitem [{\citenamefont {Gronau}\ and\ \citenamefont
  {Rosner}(2015)}]{Gronau:2015gha}%
  \BibitemOpen
  \bibfield  {author} {\bibinfo {author} {\bibfnamefont {M.}~\bibnamefont
  {Gronau}}\ and\ \bibinfo {author} {\bibfnamefont {J.~L.}\ \bibnamefont
  {Rosner}},\ }\bibfield  {title} {\bibinfo {title} {{Triple product
  asymmmetries in $\Lambda_b$ and $\Xi_b$ decays}},\ }\href
  {https://doi.org/10.1016/j.physletb.2015.07.060} {\bibfield  {journal}
  {\bibinfo  {journal} {Phys. Lett. B}\ }\textbf {\bibinfo {volume} {749}},\
  \bibinfo {pages} {104} (\bibinfo {year} {2015})},\ \Eprint
  {https://arxiv.org/abs/1506.01346} {arXiv:1506.01346 [hep-ph]} \BibitemShut
  {NoStop}%
\bibitem [{\citenamefont {Durieux}(2016)}]{Durieux:2016nqr}%
  \BibitemOpen
  \bibfield  {author} {\bibinfo {author} {\bibfnamefont {G.}~\bibnamefont
  {Durieux}},\ }\bibfield  {title} {\bibinfo {title} {{CP violation in
  multibody decays of beauty baryons}},\ }\href
  {https://doi.org/10.1007/JHEP10(2016)005} {\bibfield  {journal} {\bibinfo
  {journal} {JHEP}\ }\textbf {\bibinfo {volume} {10}},\ \bibinfo {pages}
  {005}},\ \Eprint {https://arxiv.org/abs/1608.03288} {arXiv:1608.03288
  [hep-ph]} \BibitemShut {NoStop}%
\bibitem [{\citenamefont {Shi}\ \emph {et~al.}(2019)\citenamefont {Shi},
  \citenamefont {Kang}, \citenamefont {Bigi}, \citenamefont {Wang},\ and\
  \citenamefont {Peng}}]{Shi:2019vus}%
  \BibitemOpen
  \bibfield  {author} {\bibinfo {author} {\bibfnamefont {X.-D.}\ \bibnamefont
  {Shi}}, \bibinfo {author} {\bibfnamefont {X.-W.}\ \bibnamefont {Kang}},
  \bibinfo {author} {\bibfnamefont {I.}~\bibnamefont {Bigi}}, \bibinfo {author}
  {\bibfnamefont {W.-P.}\ \bibnamefont {Wang}},\ and\ \bibinfo {author}
  {\bibfnamefont {H.-P.}\ \bibnamefont {Peng}},\ }\bibfield  {title} {\bibinfo
  {title} {{Prospects for CP and P violation in $\Lambda_{c}^+$ decays at Super
  Tau Charm Facility}},\ }\href {https://doi.org/10.1103/PhysRevD.100.113002}
  {\bibfield  {journal} {\bibinfo  {journal} {Phys. Rev. D}\ }\textbf {\bibinfo
  {volume} {100}},\ \bibinfo {pages} {113002} (\bibinfo {year} {2019})},\
  \Eprint {https://arxiv.org/abs/1904.12415} {arXiv:1904.12415 [hep-ph]}
  \BibitemShut {NoStop}%
\bibitem [{\citenamefont {Wang}\ \emph {et~al.}(2022)\citenamefont {Wang},
  \citenamefont {Qin},\ and\ \citenamefont {Yu}}]{Wang:2022fih}%
  \BibitemOpen
  \bibfield  {author} {\bibinfo {author} {\bibfnamefont {J.-P.}\ \bibnamefont
  {Wang}}, \bibinfo {author} {\bibfnamefont {Q.}~\bibnamefont {Qin}},\ and\
  \bibinfo {author} {\bibfnamefont {F.-S.}\ \bibnamefont {Yu}},\ }\bibfield
  {title} {\bibinfo {title} {{CP violation induced by T-odd correlations and
  its baryonic application}},\ }\href@noop {} {\  (\bibinfo {year} {2022})},\
  \Eprint {https://arxiv.org/abs/2211.07332} {arXiv:2211.07332 [hep-ph]}
  \BibitemShut {NoStop}%
\bibitem [{\citenamefont {Cabibbo}\ and\ \citenamefont
  {Maksymowicz}(1965)}]{Cabibbo:1965zzb}%
  \BibitemOpen
  \bibfield  {author} {\bibinfo {author} {\bibfnamefont {N.}~\bibnamefont
  {Cabibbo}}\ and\ \bibinfo {author} {\bibfnamefont {A.}~\bibnamefont
  {Maksymowicz}},\ }\bibfield  {title} {\bibinfo {title} {{Angular Correlations
  in $K_{e4}$ Decays and Determination of Low-Energy $\pi-\pi$ Phase Shifts}},\
  }\href {https://doi.org/10.1103/PhysRev.137.B438} {\bibfield  {journal}
  {\bibinfo  {journal} {Phys. Rev.}\ }\textbf {\bibinfo {volume} {137}},\
  \bibinfo {pages} {B438} (\bibinfo {year} {1965})},\ \bibinfo {note}
  {[Erratum: Phys.Rev. 168, 1926 (1968)]}\BibitemShut {NoStop}%
\bibitem [{\citenamefont {Qi}\ \emph {et~al.}(2025)\citenamefont {Qi},
  \citenamefont {Wang}, \citenamefont {Zhang},\ and\ \citenamefont
  {Guo}}]{Qi:2025zna}%
  \BibitemOpen
  \bibfield  {author} {\bibinfo {author} {\bibfnamefont {J.-J.}\ \bibnamefont
  {Qi}}, \bibinfo {author} {\bibfnamefont {Z.-Y.}\ \bibnamefont {Wang}},
  \bibinfo {author} {\bibfnamefont {Z.-H.}\ \bibnamefont {Zhang}},\ and\
  \bibinfo {author} {\bibfnamefont {X.-H.}\ \bibnamefont {Guo}},\ }\bibfield
  {title} {\bibinfo {title} {{Normalization of partial wave CP asymmetries in
  three-body decays of heavy hadrons}},\ }\href@noop {} {\  (\bibinfo {year}
  {2025})},\ \Eprint {https://arxiv.org/abs/2511.12445} {arXiv:2511.12445
  [hep-ph]} \BibitemShut {NoStop}%
\bibitem [{\citenamefont {Aaij}\ \emph {et~al.}(2023)\citenamefont {Aaij} \emph
  {et~al.}}]{LHCb:2022orj}%
  \BibitemOpen
  \bibfield  {author} {\bibinfo {author} {\bibfnamefont {R.}~\bibnamefont
  {Aaij}} \emph {et~al.} (\bibinfo {collaboration} {LHCb}),\ }\bibfield
  {title} {\bibinfo {title} {{Search for CP violation using $\hat{T}$-odd
  correlations in $B^0\to p\bar{p}K^+\pi^-$ decays}},\ }\href
  {https://doi.org/10.1103/PhysRevD.108.032007} {\bibfield  {journal} {\bibinfo
   {journal} {Phys. Rev. D}\ }\textbf {\bibinfo {volume} {108}},\ \bibinfo
  {pages} {032007} (\bibinfo {year} {2023})},\ \Eprint
  {https://arxiv.org/abs/2205.08973} {arXiv:2205.08973 [hep-ex]} \BibitemShut
  {NoStop}%
\bibitem [{\citenamefont {Wang}\ \emph {et~al.}(2004)\citenamefont {Wang} \emph
  {et~al.}}]{Belle:2003pwf}%
  \BibitemOpen
  \bibfield  {author} {\bibinfo {author} {\bibfnamefont {M.~Z.}\ \bibnamefont
  {Wang}} \emph {et~al.} (\bibinfo {collaboration} {Belle}),\ }\bibfield
  {title} {\bibinfo {title} {{Observation of $B^+ \to p \bar{p} \pi^+$, $B^0\to
  p \bar{p} K^0$, and $B^+\to p\bar{p} K^{\ast+}$}},\ }\href
  {https://doi.org/10.1103/PhysRevLett.92.131801} {\bibfield  {journal}
  {\bibinfo  {journal} {Phys. Rev. Lett.}\ }\textbf {\bibinfo {volume} {92}},\
  \bibinfo {pages} {131801} (\bibinfo {year} {2004})},\ \Eprint
  {https://arxiv.org/abs/hep-ex/0310018} {arXiv:hep-ex/0310018} \BibitemShut
  {NoStop}%
\bibitem [{\citenamefont {Suzuki}(2007)}]{Suzuki:2006nn}%
  \BibitemOpen
  \bibfield  {author} {\bibinfo {author} {\bibfnamefont {M.}~\bibnamefont
  {Suzuki}},\ }\bibfield  {title} {\bibinfo {title} {{Partial waves of
  baryon-antibaryon in three-body B meson decay}},\ }\href
  {https://doi.org/10.1088/0954-3899/34/2/009} {\bibfield  {journal} {\bibinfo
  {journal} {J. Phys. G}\ }\textbf {\bibinfo {volume} {34}},\ \bibinfo {pages}
  {283} (\bibinfo {year} {2007})},\ \Eprint
  {https://arxiv.org/abs/hep-ph/0609133} {arXiv:hep-ph/0609133} \BibitemShut
  {NoStop}%
\bibitem [{\citenamefont {Geng}\ and\ \citenamefont
  {Hsiao}(2006)}]{Geng:2006wz}%
  \BibitemOpen
  \bibfield  {author} {\bibinfo {author} {\bibfnamefont {C.~Q.}\ \bibnamefont
  {Geng}}\ and\ \bibinfo {author} {\bibfnamefont {Y.~K.}\ \bibnamefont
  {Hsiao}},\ }\bibfield  {title} {\bibinfo {title} {{Angular distributions in
  three-body baryonic B decays}},\ }\href
  {https://doi.org/10.1103/PhysRevD.74.094023} {\bibfield  {journal} {\bibinfo
  {journal} {Phys. Rev. D}\ }\textbf {\bibinfo {volume} {74}},\ \bibinfo
  {pages} {094023} (\bibinfo {year} {2006})},\ \Eprint
  {https://arxiv.org/abs/hep-ph/0606141} {arXiv:hep-ph/0606141} \BibitemShut
  {NoStop}%
\bibitem [{\citenamefont {Qi}\ \emph {et~al.}(2024)\citenamefont {Qi},
  \citenamefont {Yang},\ and\ \citenamefont {Zhang}}]{Qi:2024zau}%
  \BibitemOpen
  \bibfield  {author} {\bibinfo {author} {\bibfnamefont {J.-J.}\ \bibnamefont
  {Qi}}, \bibinfo {author} {\bibfnamefont {J.-Y.}\ \bibnamefont {Yang}},\ and\
  \bibinfo {author} {\bibfnamefont {Z.-H.}\ \bibnamefont {Zhang}},\ }\bibfield
  {title} {\bibinfo {title} {{CP asymmetries corresponding to the imaginary
  parts of the interference terms in cascade decays of heavy hadrons}},\ }\href
  {https://doi.org/10.1103/PhysRevD.110.L111301} {\bibfield  {journal}
  {\bibinfo  {journal} {Phys. Rev. D}\ }\textbf {\bibinfo {volume} {110}},\
  \bibinfo {pages} {L111301} (\bibinfo {year} {2024})},\ \Eprint
  {https://arxiv.org/abs/2407.20586} {arXiv:2407.20586 [hep-ph]} \BibitemShut
  {NoStop}%
\bibitem [{\citenamefont {Zhang}(2023)}]{Zhang:2022iye}%
  \BibitemOpen
  \bibfield  {author} {\bibinfo {author} {\bibfnamefont {Z.-H.}\ \bibnamefont
  {Zhang}},\ }\bibfield  {title} {\bibinfo {title} {{Searching for CP violation
  through two-dimensional angular distributions in four-body decays of bottom
  and charmed baryons}},\ }\href {https://doi.org/10.1103/PhysRevD.107.L011301}
  {\bibfield  {journal} {\bibinfo  {journal} {Phys. Rev. D}\ }\textbf {\bibinfo
  {volume} {107}},\ \bibinfo {pages} {L011301} (\bibinfo {year} {2023})},\
  \Eprint {https://arxiv.org/abs/2209.13196} {arXiv:2209.13196 [hep-ph]}
  \BibitemShut {NoStop}%
\bibitem [{\citenamefont {Chen}\ \emph {et~al.}(2008)\citenamefont {Chen} \emph
  {et~al.}}]{Belle:2008zkc}%
  \BibitemOpen
  \bibfield  {author} {\bibinfo {author} {\bibfnamefont {J.~H.}\ \bibnamefont
  {Chen}} \emph {et~al.} (\bibinfo {collaboration} {Belle}),\ }\bibfield
  {title} {\bibinfo {title} {{Observation of $B^0 \to p \bar{p} K^{\ast 0}$
  with a large $K^{\ast 0}$ polarization}},\ }\href
  {https://doi.org/10.1103/PhysRevLett.100.251801} {\bibfield  {journal}
  {\bibinfo  {journal} {Phys. Rev. Lett.}\ }\textbf {\bibinfo {volume} {100}},\
  \bibinfo {pages} {251801} (\bibinfo {year} {2008})},\ \Eprint
  {https://arxiv.org/abs/0802.0336} {arXiv:0802.0336 [hep-ex]} \BibitemShut
  {NoStop}%
\bibitem [{\citenamefont {Williams}(2011)}]{Williams:2011cd}%
  \BibitemOpen
  \bibfield  {author} {\bibinfo {author} {\bibfnamefont {M.}~\bibnamefont
  {Williams}},\ }\bibfield  {title} {\bibinfo {title} {{Observing CP Violation
  in Many-Body Decays}},\ }\href {https://doi.org/10.1103/PhysRevD.84.054015}
  {\bibfield  {journal} {\bibinfo  {journal} {Phys. Rev. D}\ }\textbf {\bibinfo
  {volume} {84}},\ \bibinfo {pages} {054015} (\bibinfo {year} {2011})},\
  \Eprint {https://arxiv.org/abs/1105.5338} {arXiv:1105.5338 [hep-ex]}
  \BibitemShut {NoStop}%
\bibitem [{\citenamefont {Aaij}\ \emph {et~al.}(2017)\citenamefont {Aaij} \emph
  {et~al.}}]{LHCb:2016qbq}%
  \BibitemOpen
  \bibfield  {author} {\bibinfo {author} {\bibfnamefont {R.}~\bibnamefont
  {Aaij}} \emph {et~al.} (\bibinfo {collaboration} {LHCb}),\ }\bibfield
  {title} {\bibinfo {title} {{Search for $CP$ violation in the phase space of
  $D^0\rightarrow\pi^+\pi^-\pi^+\pi^-$ decays}},\ }\href
  {https://doi.org/10.1016/j.physletb.2017.03.062} {\bibfield  {journal}
  {\bibinfo  {journal} {Phys. Lett. B}\ }\textbf {\bibinfo {volume} {769}},\
  \bibinfo {pages} {345} (\bibinfo {year} {2017})},\ \Eprint
  {https://arxiv.org/abs/1612.03207} {arXiv:1612.03207 [hep-ex]} \BibitemShut
  {NoStop}%
\bibitem [{\citenamefont {Parkes}\ \emph {et~al.}(2017)\citenamefont {Parkes},
  \citenamefont {Chen}, \citenamefont {Brodzicka}, \citenamefont {Gersabeck},
  \citenamefont {Dujany},\ and\ \citenamefont {Barter}}]{Parkes:2016yie}%
  \BibitemOpen
  \bibfield  {author} {\bibinfo {author} {\bibfnamefont {C.}~\bibnamefont
  {Parkes}}, \bibinfo {author} {\bibfnamefont {S.}~\bibnamefont {Chen}},
  \bibinfo {author} {\bibfnamefont {J.}~\bibnamefont {Brodzicka}}, \bibinfo
  {author} {\bibfnamefont {M.}~\bibnamefont {Gersabeck}}, \bibinfo {author}
  {\bibfnamefont {G.}~\bibnamefont {Dujany}},\ and\ \bibinfo {author}
  {\bibfnamefont {W.}~\bibnamefont {Barter}},\ }\bibfield  {title} {\bibinfo
  {title} {{On model-independent searches for direct CP violation in multi-body
  decays}},\ }\href {https://doi.org/10.1088/1361-6471/aa75a5} {\bibfield
  {journal} {\bibinfo  {journal} {J. Phys. G}\ }\textbf {\bibinfo {volume}
  {44}},\ \bibinfo {pages} {085001} (\bibinfo {year} {2017})},\ \Eprint
  {https://arxiv.org/abs/1612.04705} {arXiv:1612.04705 [hep-ex]} \BibitemShut
  {NoStop}%
\end{thebibliography}%

\section*{Appendix}
{\it Discussions on polarized mother hadron ---}
In general, the DAS for $H_Q \to a(\to 12) b(\to 34)$ can be expressed as 
	\[{\left| \mathcal{A}\right|^2} \propto \sum_{\sigma_a\sigma_{a'}\sigma_b\sigma_{b'}}\sum_{j_aj_b}\gamma^{j_aj_b}_{\sigma_{a}\sigma_{b}\sigma_{a'}\sigma_{b'}} \Omega^{j_aj_b}_{\sigma_{a}\sigma_{b}\sigma_{a'}\sigma_{b'}},\]
where $\gamma^{j_aj_b}_{\sigma_{a}\sigma_{b}\sigma_{a'}\sigma_{b'}}$ and $\Omega^{j_aj_b}_{\sigma_{a}\sigma_{b}\sigma_{a'}\sigma_{b'}}$ are respectively the dynamical and kinematical factors, 
$\sigma_{a^{(\prime)}}$ and $\sigma_{b^{(\prime)}}$ are respectively the helicity indices for $a^{(\prime)}$ and $b^{(\prime)}$.
The kinematical factor takes the form $\Omega^{j_aj_b}_{\sigma_{a}\sigma_{b}\sigma_{a'}\sigma_{b'}} \!\!\equiv \! P_{\sigma_{ab},\sigma_{a'b'}} \!(\theta_{H_Q}) d^{j_a}_{\sigma_{a'a},0} \!({\theta_a})d^{j_b}_{\sigma_{b'b},0} \!({\theta_b}) e^{i(\bar{\sigma}\varphi+\hat{\sigma}\phi)}$,
where $P$ is the polarization matrix of the hadron $H_Q$, 
 $\bar{\sigma}=({\sigma_{a'a}+\sigma_{b'b}})/{2}$,  and  $\hat{\sigma}=({\sigma_{a'a}-\sigma_{b'b}})/{2}$, with
$\sigma_{xy}=\sigma_{x}-\sigma_y$ $(x,y=a^{(\prime)},b^{(\prime)})$.
The dynamical factor takes the form
$\gamma^{j_aj_b}_{\sigma_{a}\sigma_{b}\sigma_{a'}\sigma_{b'}}=\sum_{a,a'=a_k \atop b,b'=b_m}
{{w}^{(ab,a'b')j_aj_b}_{\sigma_{a}\sigma_{b}\sigma_{a'}\sigma_{b'}}\mathcal{G}^{(aa')}_{j_a}\mathcal{G}^{(bb')}_{j_b}}$,
where
	${w}^{(ab,a'b'){j_aj_b}}_{\sigma_{a}\sigma_{b}\sigma_{a'}\sigma_{b'}} 
	=(-)^{\sigma_{a}-s_a+\sigma_{b}-s_b}
	C{{\mathstrut}^{\,\,\,\,s_{a}}_{-\sigma_a} {\mathstrut}^{s_{a'}}_{\sigma_{a'}}{\mathstrut}^{j_a}_{\sigma_{a^\prime\!a}}} 
	C{{\mathstrut}^{\,\,\,\,s_{b}}_{-\sigma_b} {\mathstrut}^{s_{b'}}_{\sigma_{b'}} {\mathstrut}^{j_b}_{\sigma_{b'\!b}}}
	\mathcal{F}^{H\to ab}_{\sigma_{a}\sigma_{b}}\mathcal{F}^{H\to a'b' \ast}_{\sigma_{a'}\sigma_{b'}}$,
and 
	$\mathcal{G}^{(aa')}_{j_a} \!\!=\! \frac{1}{\mathcal{I}_a\mathcal{I}_{a'}^\ast} \!\! \sum_{\lambda_1\lambda_2}(-)^{s_a-\lambda_{12}} C{{\mathstrut}^{\,\,\,\,s_a}_{-\!{\lambda}_{12}}}{{\mathstrut}^{s_{a'}}_{\lambda_{12}}}{\mathstrut}^{j_a}_{0} \mathcal{F}^{a\to 12}_{\lambda_1\lambda_2}\mathcal{F}^{a'\to 12\ast}_{\lambda_1\lambda_2}$,
	$\mathcal{G}^{(bb')}_{j_b} \!\!=\! \frac{1}{\mathcal{I}_b\mathcal{I}_{b'}^\ast} \!\! \sum_{\lambda_3\lambda_4}(-)^{s_b-\lambda_{34}} C {{}^{\,\,\,\,s_b}_{-\lambda_{34}} {}^{s_{b'}}_{\lambda_{34}} {\mathstrut}^{j_b}_{0}}
	\mathcal{F}^{b\to34}_{\lambda_3\lambda_4}\mathcal{F}^{b'\to34\ast}_{\lambda_3\lambda_4}$,
where all the $\mathcal{F}$'s are the helicity decay amplitudes, 
$\lambda_k$ is the helicity of the particle $k$ in the final state, with $\lambda_{kl}=\lambda_k-\lambda_l$, $s_{a^{(\prime)}}$ and $s_{b^{(\prime)}}$ are the spins of ${a^{(\prime)}}$ and ${b^{(\prime)}}$, respectively, $\mathcal{I}_{r}=m^2-m^2_r+im_r\Gamma_r$, with $m$ being $m_{12}$ or $m_{34}$, {\it i.e.}, the inverse of the Breit-Wigner factor if $r$ is a resonance, and $1/\mathcal{I}_{r}$ can be a smooth function of $m^2$ (or simply a constant with respect to $m^2$) if $r$ is not a resonance, and $C{{}^{j_1}_{m_1}{}^{j_2}_{m_2}{}^{j}_{m}}$ are the Clebsch-Gordan coefficients.

For polarized $H_Q$ of spin-$J$, 
the polarization matrix can be generally expressed as $P=\frac{1}{2J+1}\big(1+\sum_{k=1}^{J}\sum_{q=-k}^{k}C_{kq}\mathcal{O}_{q}^{(k)}\big)$, where $\mathcal{O}_{q}^{(k)}$ are the covariant spin operators, $C_{kq}$ are the coefficients that depend on the polarization state of the hadron $H_Q$.
For example, for spin-half baryons such as $\Lambda_b^0$ and $\Lambda_c^+$, the polarization matrix takes the form
	$P(\theta_{H_Q})=\frac{1}{2}\left[1+\left(\begin{array}{c c} \cos\theta_{H_Q} & -\sin\theta_{H_Q}\\ -\sin\theta_{H_Q} & -\cos\theta_{H_Q}
	\end{array}\right)P_z\right]$,
where $P_z$ is the polarization of $H_Q$, which is chosen to be along the $z$ axis.
For the unpolarized $H_Q$, one just needs to make the replacement $P_{\sigma_{ab},\sigma_{a'b'}} \rightarrow\delta_{\sigma_{ab},\sigma_{a'b'}}/(2J+1)$.
Consequently, one has $\sigma_{ab}=\sigma_{a'b'}$, hence $\sigma_{a'a}=\sigma_{b'b}$, which allow us to denote $\sigma_{a'a}=\sigma_{b'b}\equiv\sigma$.
Then the kinematical factors become the following compact form
\[\Omega^{j_aj_b}_{\sigma}= d^{j_a}_{\sigma,0}({\theta_a})d^{j_b}_{\sigma,0}({\theta_b}) e^{i\sigma\varphi}=\Psi_{\sigma}^{j_aj_b}+i \Phi_{\sigma}^{j_aj_b}.\]
One can see that {\it $\varphi$ are correlated with the two helicity angles $\theta_a$ and $\theta_b$, and {\it always} come together with $\Phi_{\sigma}^{j_aj_b}$ and $\Psi_{\sigma}^{j_aj_b}$ for unpolarized $H_Q$}.
The first few angular correlations are presented in TABLE \ref{Tab:AnCo}.
The correlations between the kinematical angles in Eqs. (\ref{Eq:Phi}) and (\ref{Eq:Psi}) have important consequences.
They allow us to re-express the DAS into Eq. (\ref{eq:AngDisCom})
where the compact dynamical factor $\gamma^{j_aj_b}_{\sigma}$ takes the form 
	$\gamma^{j_aj_b}_{\sigma}=\sideset{}{'}\sum_{\sigma_a\sigma_{a'}\sigma_b\sigma_{b'}} \gamma^{j_aj_b}_{\sigma_{a}\sigma_{b}\sigma_{a'}\sigma_{b'}}$,
where the prime in the superscript of the summation symbol indicates that this summation is constrained by $-s_{H_Q}\leqslant \sigma_{ab}=\sigma_{a'b'}\leqslant s_{H_Q}$ and $\sigma_{aa'}=\sigma_{bb'}=\sigma$.

The non-zero off diagonal elements, which are proportional to $-P_z\sin\theta_{H_Q}$ for spin-half $H_Q$, 
indicate that there is no more constraint on the equality between $\sigma_{a'a}$ and $\sigma_{b'b}$. 
However, the difference between them is at most 1:  $\sigma_{a'a}=\sigma_{b'b}\pm1$, which is in fact restricted by the spin of $H_Q$.
Hence, the terms corresponding to the polarization of $H_Q$ take the from
 $d^{j_a}_{\bar{\sigma}\pm\frac{1}{2},0}(\theta_a)d^{j_b}_{\bar{\sigma}\mp\frac{1}{2},0}(\theta_b)\cos(\bar{\sigma}\varphi\pm\frac{1}{2}\phi)$ and $d^{j_a}_{\bar{\sigma}\pm\frac{1}{2},0}(\theta_a)d^{j_b}_{\bar{\sigma}\mp\frac{1}{2},0}(\theta_b)\sin(\bar{\sigma}\varphi\pm\frac{1}{2}\phi)$.
 These are the decay angular correlations which should be taken into account for polarized $H_Q$.

\begin{table*}
  \centering
\begin{tabular}{|c|l|l|l|}
  \hline
  \diagbox[height=20pt,innerwidth=20pt]{${j_a}$}{${j_b}$}
    & \multicolumn{1}{c|}{0} & \multicolumn{1}{c|}{1} & \multicolumn{1}{c|}{2}  \\
    \hline
  0 & $\Psi^{00}_0=1$ & $\Psi^{01}_0=d^{1}_{00}(\theta_b)=c_{\theta_b}$ & $\Psi^{02}_0=d^{2}_{00}(\theta_b)=\frac{1}{2}(3c_{\theta_b}^2 \!\!-\!1)$   \\
    \hline
 \multirow{3}{*}{1} & $\Psi^{10}_0=d^{1}_{00}(\theta_a)=c_{\theta_a}$  & $\Psi^{11}_0=d^{1}_{00}(\theta_a)d^{1}_{00}(\theta_b)=c_{\theta_a}c_{\theta_b}$ & $\Psi^{12}_0=d^{1}_{00}(\theta_a)d^{2}_{00}(\theta_b)=\frac{1}{2}c_{\theta_a}(3c_{\theta_b}^2 \!\!-\! 1)$   \\
   &  & $\Psi^{11}_1=d^{1}_{10}(\theta_a)d^{1}_{10}(\theta_b)c_\varphi=s_{\theta_a}s_{\theta_b}c_\varphi$ &  $\Psi^{12}_1=d^{1}_{10}(\theta_a)d^{2}_{10}(\theta_b)c_\varphi=s_{\theta_a}s_{\theta_b}c_{\theta_b}c_\varphi$   \\
   &  & $\Phi^{11}_1=d^{1}_{10}(\theta_a)d^{1}_{10}(\theta_b)s_\varphi=s_{\theta_a}s_{\theta_b}s_\varphi$ & $\Phi^{12}_1=d^{1}_{10}(\theta_a)d^{2}_{10}(\theta_b)s_\varphi=s_{\theta_a}s_{\theta_b}c_{\theta_b}s_\varphi$ \\
     \hline
 \multirow{5}{*}{2} &  $\Psi^{20}_0=d^{2}_{00}(\theta_a)=\frac{1}{2}(3c_{\theta_a}^2 \!\!-\! 1)$ & $\Psi^{21}_0=d^{2}_{00}(\theta_a)d^{1}_{00}(\theta_b)=\frac{1}{2}(3c_{\theta_a}^2 \!\!-\! 1)c_{\theta_b}$ & $\Psi^{22}_0=d^{2}_{00}(\theta_a)d^{2}_{00}(\theta_b)=\frac{1}{4}(3c_{\theta_a}^2 \!\!-\! 1)(3c_{\theta_b}^2 \!\!-\! 1)$ \\
   &   & $\Psi^{21}_1=d^{2}_{10}(\theta_a)d^{1}_{10}(\theta_b)c_\varphi=s_{\theta_a}c_{\theta_a}s_{\theta_b}c_\varphi$ &  $\Psi^{22}_1=d^{2}_{10}(\theta_a)d^{2}_{10}(\theta_b)c_\varphi =s_{\theta_a}c_{\theta_a}s_{\theta_b}c_{\theta_b}c_{\varphi}$ \\
   &   & $\Phi^{21}_1=d^{2}_{10}(\theta_a)d^{1}_{10}(\theta_b)s_\varphi =s_{\theta_a}c_{\theta_a}s_{\theta_b}s_\varphi$ & $\Phi^{22}_1=d^{2}_{10}(\theta_a)d^{2}_{10}(\theta_b)s_\varphi =s_{\theta_a}c_{\theta_a}s_{\theta_b}c_{\theta_b}s_\varphi$   \\
   &   &   & $\Psi^{22}_2=d^{2}_{20}(\theta_a)d^{2}_{20}(\theta_b) c_{2\varphi}=\frac{3}{8} s_{\theta_a}^2 s_{\theta_b}^2 c_{2\varphi}$  \\
   &   &   & $\Phi^{22}_2=d^{2}_{20}(\theta_a)d^{2}_{20}(\theta_b) s_{2\varphi} =\frac{3}{8} s_{\theta_a}^2 s_{\theta_b}^2 s_{2\varphi}$   \\
  \hline
\end{tabular}
 \caption{The first few angular correlations. Abbreviations similar to those in TABLE \ref{Tab:B0Data} are adopted.} \label{Tab:AnCo}
\end{table*}

{\it Extracting the event yields---}
We present here the event yields extracted from the data of TPAs of LHCb in Ref. \cite{LHCb:2022orj}.
As is mentioned in the main text, we focus on the phase space region when the invariant mass of the $p\bar{p}$ system is small, and the invariant mass of the $K^+\pi^-$ system is around that of $K^*(892)$.
This means that what we will concern is the phase space regions of Region 0 to Region 7 in Scheme A, and Region 0 to Region 15 in Scheme B of the bin division in Ref. \cite{LHCb:2022orj}. 
Note that Scheme B in fact further divides each region in Scheme A into two regions according to the sign of $m_{K\pi}^2-m_{K^\ast}^2$.

In TABLEs \ref{Tab:B0Data} and \ref{Tab:B0barData}, we present the measured TPAs in these regions that we are concerning.
With these measured TPAs, we are able to extract the event yields for $\sin\varphi>0$ and $\sin\varphi<0$ for each region according to 
		$A_{T,i}\pm\sigma_{A_{T,i}}=\frac{N_{i,s_{\varphi>0}}-N_{i,s_{\varphi}<0}}{N_{i,s_{\varphi>0}}+N_{i,s_{\varphi}<0}}\pm{\frac{k}{\sqrt{N_{i,s_{\varphi>0}}+N_{i,s_{\varphi}<0}}}}$,
where we have introduced a scale factor $k$ which expresses the uncertainties $\sigma_{A_{T,i}}$.
The introduction of the universal scale factor $k$ is based on the observation that there is a clear alignment between the background subtracted yields (projected to the kinematic variables $\theta_{p\overline{p}}$, $\theta_{K\pi}$, and $\varphi$) and the corresponding errors, which can be read off from Figure 5, $N=(\sigma_N/k)^2$, where $k$ is fitted to be $k=1.55$.
We assume that the universal scale factor $k$ also apply to the bins in TABLEs \ref{Tab:B0Data} and \ref{Tab:B0barData}.
	
There are fluctuations making the real yields deviate from the perfect behavior $N=(\sigma_N/k)^2$. 
These potential bias are estimated to be roughly the same as the corresponding statistical uncertainties based on the analysis of the data in Figure 5 of Ref. \cite{LHCb:2022orj}. 
However, we did not include this in the main text. That is because the uncertainties caused by the fitting only enter in our procedure of extracting the yields from the data. In the real experiment, such a procedure is not needed.

The extracted yields are also presented in these two Tables.
In each row, the sum of the two event yields in Scheme B should approximately equal to the event yield in Scheme A.
It seems that there are two typos in the data of TPAs presented in Region 9 of Scheme B for both of the $B^0$ and $\overline{B^0}$ cases. 
Namely, the statistical errors are about two times smaller, so we have made corresponding modifications. 
It should be pointed out this modification does not affect the basic conclusion here. For example, the angular-correlated CPA $\tilde{\Psi}^{01}_0$ in TABLE \ref{Tab:ACPX} would be $(9.5\pm1.87\pm0.19)\%$, if the data in Region 9 of Scheme B had not been modified.

\begin{table*}[b]
	\centering
	\begin{tabular}{|||c|c|||c|c|c|||c|c|c||c|c|c|||}
		\hline
		\multicolumn{2}{|||c|||}{${B^0}$}&\multicolumn{3}{c|||}{Scheme A}&\multicolumn{6}{c|||}{Scheme B}\\
		\hline
		\multirow{2}{*}{\shortstack{sign of\\ ${c}_{\theta_a}$ ${c}_{\theta_b}$ ${c}_\varphi$}} &  \multirow{2}{*}{\shortstack{sign of\\  ${s}_\varphi$}} & \multirow{2}{*}{region} & \multirow{2}{*}{$A_{\hat{T}}$} &  \multirow{2}{*}{yields} & \multicolumn{3}{c||}{$m^2_{K\pi}-m_{K^\ast}^2<0$}
		& \multicolumn{3}{c|||}{$m^2_{K\pi}-m_{K^\ast}^2>0$}  \\
		\cline{6-11}
		&  &  & & & region & $A_{\hat{T}}$ & yields & region & $A_{\hat{T}}$ & yields  \\
		\hline
		\multirow{2}{*}{$--+$}& $+$ & \multirow{2}{*}{0} & \multirow{2}{*}{$-16.5\pm10.1$}& 98 & \multirow{2}{*}{0} & \multirow{2}{*}{$-26.7\pm17.8$} & 28 & \multirow{2}{*}{8} & \multirow{2}{*}{$-5.1\pm12.8$} & 70  \\
		\cline{2-2} \cline{5-5} \cline{8-8} \cline{11-11}
		& $-$ & &  & 137 & & & 48 & & & {77} \\
		\hline
		\multirow{2}{*}{$---$} & $+$ & \multirow{2}{*}{1} & \multirow{2}{*}{$6.1\pm9.2$}&  150 & \multirow{2}{*}{1} & \multirow{2}{*}{$5.4\pm15.8$} & 51 & \multirow{2}{*}{9} & \multirow{2}{*}{$6.6\pm{11.6}$} & 95 \\
		\cline{2-2}\cline{5-5} \cline{8-8} \cline{11-11}
		& $-$ & &  & 133 & & & 45 & & & 83  \\
		\hline
		\multirow{2}{*}{$-++$} & $+$ & \multirow{2}{*}{2} & \multirow{2}{*}{$-1.2\pm7.0$} & 242 & \multirow{2}{*}{2} & \multirow{2}{*}{$-7.3\pm11.1$} & 90 & \multirow{2}{*}{10} & \multirow{2}{*}{$0.7\pm9.0$} & 149 \\
		\cline{2-2}\cline{5-5} \cline{8-8} \cline{11-11}
		& $-$ & &  & 248 & & & 105 & & & 147 \\
		\hline
		\multirow{2}{*}{$-+-$} & $+$ & \multirow{2}{*}{3} & \multirow{2}{*}{$25.3\pm7.2$} & 290 & \multirow{2}{*}{3} & \multirow{2}{*}{$15.4\pm12.8$} & 85 & \multirow{2}{*}{11} & \multirow{2}{*}{$30.9\pm8.7$} & 208 \\
		\cline{2-2}\cline{5-5} \cline{8-8} \cline{11-11}
		& $-$ & &  & 173 & & & 62 & & & 110 \\
		\hline
		\multirow{2}{*}{$+-+$} & $+$ & \multirow{2}{*}{4} & \multirow{2}{*}{$7.8\pm11.1$} & 105 & \multirow{2}{*}{4} & \multirow{2}{*}{$-21.9\pm13.9$} & 49 & \multirow{2}{*}{12} & \multirow{2}{*}{$38.4\pm16.8$} & 59 \\
		\cline{2-2}\cline{5-5} \cline{8-8} \cline{11-11}
		& $-$ & &  & 90 & & & 76 & & & {26}\\
		\hline
		\multirow{2}{*}{$+--$} & $+$ & \multirow{2}{*}{5} & \multirow{2}{*}{$2.9\pm8.3$} & 179 & \multirow{2}{*}{5} & \multirow{2}{*}{$-13.4\pm13.9$} & 54 & \multirow{2}{*}{13} & \multirow{2}{*}{$11.6\pm10.2$} & 129 \\
		\cline{2-2}\cline{5-5} \cline{8-8} \cline{11-11}
		& $-$ & &  & 169 & & & 70 & & & 102 \\
		\hline
		\multirow{2}{*}{$+++$} & $+$ & \multirow{2}{*}{6} & \multirow{2}{*}{$-22.8\pm7.4$} & 169 & \multirow{2}{*}{6} & \multirow{2}{*}{$-19.3\pm10.4$} & 90 & \multirow{2}{*}{14} & \multirow{2}{*}{$-24.1\pm10.5$} & 83 \\
		\cline{2-2}\cline{5-5} \cline{8-8} \cline{11-11}
		& $-$ & &  & 269 & & & 132 & & & 135 \\
		\hline
		\multirow{2}{*}{$++-$} & $+$ & \multirow{2}{*}{7} & \multirow{2}{*}{$-10.4\pm6.8$} & 233 & \multirow{2}{*}{7} & \multirow{2}{*}{$0.7\pm10.9$} & 102 & \multirow{2}{*}{15} & \multirow{2}{*}{$-18.8\pm8.6$} & 132 \\
		\cline{2-2}\cline{5-5} \cline{8-8} \cline{11-11}
		& $-$ 	& &  & 287 & & & 100 & & & 193 \\
		\hline
	\end{tabular}
	\caption{The TPAs in different regions from the data of LHCb, and the corresponding event yields extracted from the TPAs data for $B^0\to p\bar{p}K^+\pi^-$. In the table, $c_{\theta_a}$, $c_{\theta_b}$, $c_\varphi$ and $s_\varphi$ are abbreviations for $\cos\theta_a$, $\cos\theta_b$, $\cos\varphi$, and $\sin\varphi$, respectively.} \label{Tab:B0Data}
\end{table*}

\begin{table*}
	\centering
	\begin{tabular}{|||c|c|||c|c|c|||c|c|c||c|c|c|||}
		\hline
		$\overline{B^0}$&\multicolumn{4}{c|||}{Scheme A}&\multicolumn{6}{c|||}{Scheme B}\\
		\hline
		\multirow{2}{*}{\shortstack{sign of\\ ${c}_{\bar{\theta}_a}$ ${c}_{\bar{\theta}_b}$ ${c}_{\bar{\varphi}}$}} &\multirow{2}{*}{\shortstack{sign of \\ ${s}_{\bar{\varphi}}$}} & \multirow{2}{*}{region} & \multirow{2}{*}{$\bar{A}_{\hat{T}}$} &  \multirow{2}{*}{yields} & \multicolumn{3}{c||}{$m^2_{K\pi}-m_{K^\ast}^2<0$}
		& \multicolumn{3}{c|||}{$m^2_{K\pi}-m_{K^\ast}^2>0$} \\
		\cline{6-11}
		&  &  &  &  & region & $\bar{A}_{\hat{T}}$ & yields & region & $\bar{A}_{\hat{T}}$ & yields \\
		\hline
		\multirow{2}{*}{$--+$} & $-$ & \multirow{2}{*}{0} & \multirow{2}{*}{$-13.2\pm9.5$} & 115 & \multirow{2}{*}{0} & \multirow{2}{*}{$-21.9\pm12.9$} & 56 & \multirow{2}{*}{8} & \multirow{2}{*}{$-8.0\pm13.2$} & 63\\
		\cline{2-2} \cline{5-5} \cline{8-8} \cline{11-11}
		& $+$ & &  & 151 & & & 88 & & & {74}\\
		\hline
		\multirow{2}{*}{$---$} & $-$& \multirow{2}{*}{1} & \multirow{2}{*}{$3.2\pm9.8$} & 129 & \multirow{2}{*}{1} & \multirow{2}{*}{$-1.6\pm20.7$} & 28 & \multirow{2}{*}{9} & \multirow{2}{*}{$4.0\pm{11.2}$} & 99\\
		\cline{2-2} \cline{5-5} \cline{8-8} \cline{11-11}
		& $+$ & &  & 121 & & & 28 & & & 92\\
		\hline
		\multirow{2}{*}{$-++$} & $-$ & \multirow{2}{*}{2} & \multirow{2}{*}{$23.9\pm10.0$} & 149 & \multirow{2}{*}{2} & \multirow{2}{*}{$18.9\pm17.4$} & 47 & \multirow{2}{*}{10} & \multirow{2}{*}{$30.2\pm12.2$} & 105\\
		\cline{2-2} \cline{5-5} \cline{8-8} \cline{11-11}
		& $+$ & &  & 91 & & & 32 & & & 56\\
		\hline
		\multirow{2}{*}{$-+-$} & $-$ & \multirow{2}{*}{3} & \multirow{2}{*}{$3.2\pm7.8$} & 204 & \multirow{2}{*}{3} & \multirow{2}{*}{$5.0\pm13.7$} & 67 & \multirow{2}{*}{11} & \multirow{2}{*}{$0.2\pm9.4$} & 136\\
		\cline{2-2} \cline{5-5} \cline{8-8} \cline{11-11}
		& $+$ & &  & 191 & & & 61 & & & 136\\
		\hline
		\multirow{2}{*}{$+-+$}& $-$ & \multirow{2}{*}{4} & \multirow{2}{*}{$24.3\pm9.0$} & 184 & \multirow{2}{*}{4} & \multirow{2}{*}{$26.1\pm16.3$} & 57 & \multirow{2}{*}{12} & \multirow{2}{*}{$22.7\pm10.7$} & 129\\
		\cline{2-2} \cline{5-5} \cline{8-8} \cline{11-11}
		& $+$ & &  & 112 & & & 33 & & & 81\\
		\hline
		\multirow{2}{*}{$+--$} & $-$ & \multirow{2}{*}{5} & \multirow{2}{*}{$14.9\pm8.6$} & 186 & \multirow{2}{*}{5} & \multirow{2}{*}{$21.9\pm22.3$} & 29 & \multirow{2}{*}{13} & \multirow{2}{*}{$14.2\pm8.8$} & {177}\\
		\cline{2-2} \cline{5-5} \cline{8-8} \cline{11-11}
		& $+$ & &  & 138 & & & 19 & & & {133}\\
		\hline
		\multirow{2}{*}{$+++$}& $-$ & \multirow{2}{*}{6} & \multirow{2}{*}{$-4.9\pm8.6$} & 154 & \multirow{2}{*}{6} & \multirow{2}{*}{$-15.3\pm11.4$} & 78 & \multirow{2}{*}{14} & \multirow{2}{*}{$6.4\pm13.2$} & {55}\\
		\cline{2-2} \cline{5-5} \cline{8-8} \cline{11-11}
		& $+$ & &  & 170 & & & 106 & & & {48}\\
		\hline
		\multirow{2}{*}{$++-$}& $-$ & \multirow{2}{*}{7} & \multirow{2}{*}{$6.8\pm6.6$} & 294 & \multirow{2}{*}{7} & \multirow{2}{*}{$2.8\pm8.4$} & 175 & \multirow{2}{*}{15} & \multirow{2}{*}{$10.2\pm9.5$} & {147}\\
		\cline{2-2} \cline{5-5} \cline{8-8} \cline{11-11}
		& $+$& &  & 257 & & & 165 & & & {119}\\
		\hline
	\end{tabular}
	\caption{The same as TABLE \ref{Tab:B0Data} but for $\overline{B^0}\to p\bar{p}K^-\pi^+$.} \label{Tab:B0barData}
\end{table*}

\end{document}